\documentclass{aa}  

\usepackage[colorlinks=true, linkcolor = blue,
            urlcolor  = blue,
            citecolor = blue,
            anchorcolor = blue, bookmarks=false]{hyperref}
\usepackage{graphicx}
\usepackage{txfonts}
\usepackage{natbib}
\usepackage{enumitem} 
\usepackage{color}
\usepackage{threeparttable}
\usepackage{tikz,amsmath}
\usetikzlibrary{shapes.geometric, arrows}
\usepackage{ulem}
\usepackage{wrapfig}
\usepackage{rotating}
\usepackage{lscape} 
\tikzstyle{process} = [rectangle, minimum width=1.5em, minimum height=3.5em, text centered, draw=blue, fill=gray!10]
\tikzstyle{process2} = [rectangle, minimum width=1.5em, minimum height=3.5em, text centered, draw=white, fill=white]
\tikzstyle{arrow} = [thick,->,>=stealth]
\usepackage{multirow}
\usepackage{cancel}
\usepackage{ulem}
\bibpunct{(}{)}{;}{f}{}{,} 

\usepackage[colorlinks=true,citecolor=blue,urlcolor=blue]{hyperref}

\newcommand{\teffa}{$T_{\mathrm{eff}}^{\mathrm{H}\alpha}$}

\newcommand{\teff}{$T_{\mathrm{eff}}$}

\newcommand{\logg}{\mbox{log \textit{g}}}

\newcommand{\titan}{\textsc{Titans}}

\newcommand{\vcalibsub}{$v_{mic}^{Ba}$}
\newcommand{\vcalibres}{$v_{mic}^{Ba 4934}$}
\newcommand{\vlte}{$v_{mic}^{LTE}$}
\newcommand{\vnlte}{$v_{mic}^{NLTE}$}
\newcommand{\vfit}{$v_{mic}^{fit\,LTE}$}
\begin{document}

   \title{Barium isotopic ratios in metal-poor stars: calibrating the method with globular clusters
   \thanks{Based on data of the {\it Gaia}-ESO survey}
   }

   \subtitle{Paper~I: Dwarf and giant stars in  NGC~6752}

   \author{R. E. Giribaldi\inst{1}
           \and
           L. Magrini\inst{1}
           \and
           J. Schiappacasse-Ulloa\inst{1}
           \and
           S. Randich\inst{1}
           \and
           T. Merle\inst{2,3}
          }
   \institute{INAF – Osservatorio Astrofisico di Arcetri, Largo E. Fermi 5, 50125
Firenze, Italy \\
            \email{riano.escategiribaldi@inaf.it} 
            \and
            Institut d'Astronomie et d'Astrophysique, Universit\'e libre de Bruxelles, CP 226, Boulevard du Triomphe, 1050 Brussels, Belgium
            \and
            Royal Observatory of Belgium, Avenue Circulaire 3, 1180 Brussels, Belgium
            }

    \date{Received NN, 2025; Accepted NN, 2025}

 
  \abstract
{
Identifying the nucleosynthesis processes behind heavy-element enrichment in stellar atmospheres is challenging. It typically relies on comparing observed abundance-to-iron ratios with theoretical predictions relative to the Sun, but this method is prone to uncertainty due to limitations of classical 1D hydrostatic models that neglect chromospheric effects.
One promising but still underexplored approach is to measure the isotopic composition of stellar atmospheres by focusing on elements that have both slow (s)-process and rapid (r)-process contributions. While the study of total elemental abundances offers a simplified view, isotopic ratios are directly linked to the underlying nucleosynthesis processes.
}  
{Our aim is to provide a reliable method for quantifying the contributions of the s- and r-processes to barium in stellar atmospheres. This is achieved by determining barium isotopic ratios using 1D atmospheric models in combination with a carefully calibrated microturbulence, based on the comparison between subordinate and resonance Ba lines. 
}
{In this initial study, we use member stars of the globular cluster NGC~6752 to calibrate the microturbulence ($v_{mic}$) value for both subordinate and resonance barium lines across different stellar evolutionary stages. This allows us to provide a reliable estimate of $v_{mic}$ that can be applied to accurately determine barium abundances and isotopic ratios in stars ranging from the main sequence to the upper red giant branch.
}
{The $v_{mic}$ scale adapted for barium subordinate lines is consistent with that derived from 3D model atmospheres, and thus the \teff-\logg\ dependent relations of the later can be used safely. 
The $v_{mic}$ for the resonance line at $\lambda$4934~\AA\ --for the determination of the isotopic ratio-- is higher, and depends on  the equivalent width (EW). 
We provide calibrated relations between $v_{mic}$ and EW for measuring isotopic ratios.
}
{}


   \maketitle
%

\section{Introduction}
Barium is primarily synthesized in the interiors of asymptotic giant branch (AGB) stars through the slow neutron-capture process (s-process). However, numerous stars exhibiting enrichment in elements associated with the rapid neutron-capture process (r-process) have also been found to show enhanced barium abundances \citep[e.g.][]{hansen2018ApJ...858...92H,da_silva2025A&A...696A.122D}.
In this context, determining barium isotopic ratios from resonance lines offers considerable potential as a precise diagnostic tool for identifying the underlying astrophysical nucleosynthesis processes—for example, by constraining the relative contribution of the r-process across different environments.
The method has been successfully applied in dwarf stars \citep{mashonkina2006A&A...456..313M},  where its diagnostic of the r-process contribution aligns well with the results inferred from [Eu/Ba] ratios. Its application to giant stars has also been explored, though so far it has been largely limited to Carbon-Enhanced Metal-Poor (CEMP) stars by several groups \citep[e.g.][]{Meng2016A&A...593A..62M,cescutti2021A&A...654A.164C,Wen2022Univ....8..596W}.

\begin{figure}
    \centering
    \includegraphics[width=0.8\linewidth]{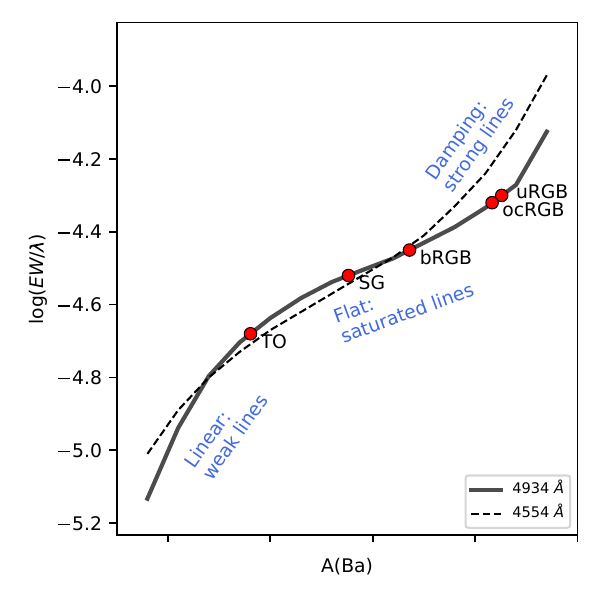}
   
    \caption{\tiny Theoretical curves of growth for the two Ba resonance lines. Equivalent widths (EW) were computed in spectra synthesised by MARCS 1D atmosphere models adopting the
    atmospheric parameters \teff\ = 5540~K, \logg\ = 2.45, [Fe/H] = $-2.00$~dex, and $v_{mic} = 1.5$~km s$^{-1}$.
    The barium abundance in the horizontal axis is neglected on purpose, as the scale depends on the abundance itself and the stellar parameters.
    The location of our stars is indicated by the red circles, where they are designated according to their evolutionary state as follows: turn-off (TO), subgiant (SG), base of the red giant branch (bRGB), over the clump RGB (ocRGB), and upper RGB (uRGB).
    }
    \label{fig:growth}
\end{figure}

Still, the limited use of Ba lines as diagnostic tools in giant stars may be attributed to their tendency to appear strong or saturated, being placed in the non-linear regime of the curve of growth. See for example Fig.~\ref{fig:growth}, which shows theoretical curves of growth of two Ba resonance lines, obtained from MARCS model atmospheres \citep{gustafson2008}. 
It is well established that one-dimensional (1D) model atmospheres struggle to accurately reproduce such strong spectral lines—not only in terms of their depth and width, but also in capturing the asymmetries between their red and blue wings, which are influenced by granulation effects \citep[e.g.][]{Asplund2000A&A...359..743A}.
Therefore, strong lines are typically avoided in abundance analyses based on 1D model atmospheres, due to their inherent limitations. 

The most robust Ba isotopic ratio measurements in very metal-poor ([Fe/H]\footnote{Here, metallicity is determined using the element Fe as a proxy. We use the iron to hydrogen fraction relative to the Sun in logarithmic scale as follows: [A/B] = $ \log{\left( \frac{N(\text{A})}{N(\text{B})} \right )_\text{Star}} - \log{\left( \frac{N(\text{A})}{N(\text{B})} \right )_\text{Sun}} $, where $N$ denotes the number abundance of a given element.} $< -2$~dex) giant stars are those of  \cite{sitnova2025}. They carefully selected field stars at the base of the Red Giant Branch (RGB) (\logg~$\lesssim$ 2.5~dex), whose Ba resonance lines remain below the plateau of the curve of growth, these have Reduced Equivalent Width (REW\footnote{log$(\text{EW}/\lambda)$, where EW is the equivalent width in \AA\ and $\lambda$ is the wavelength in \AA.}) $\lesssim -4.6$, see Fig.~\ref{fig:growth}, thus their measurements are virtually unaffected by the 1D modelling biases.

\begin{figure}
    \centering
    \includegraphics[width=0.99\linewidth]{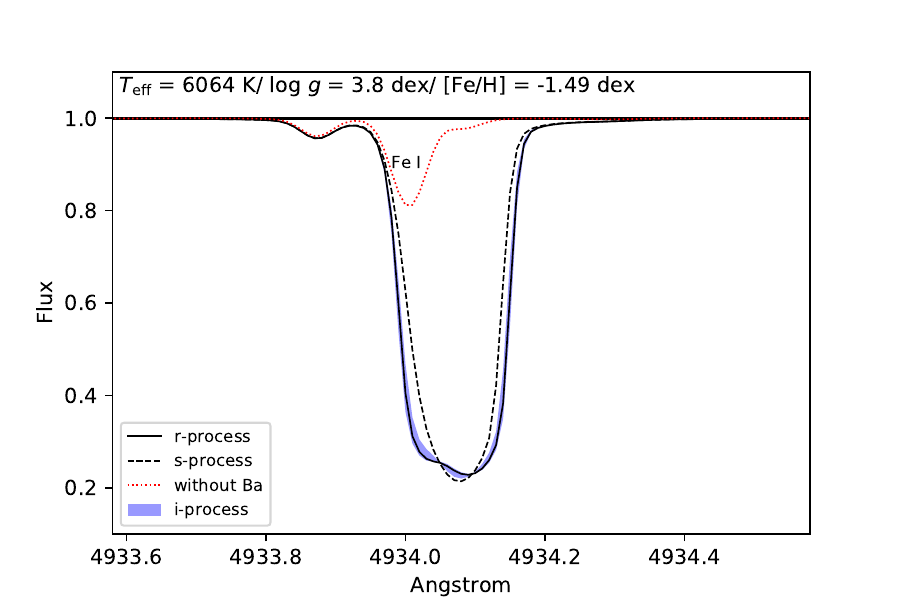}
   
    \caption{\tiny Synthetic profiles of the line at $\lambda$4934~\AA. 
    The synthesis is made with the code Turbospectrum \citep{gerber2023} and MARCS models; see details in Sect.~\ref{sec:isotopes}.
    The atmospheric parameters for the synthesis are \textbf{similar to} those of the star GES~J19102677-6003089 (TO-1) in Table~\ref{tab:atmo_param}. The spectrum has R = 493000 and infinite S/N.
    The solid line (r-process), dashed line (s-process), and blue shade (i-process) profiles are synthesised with the isotopic ratios in Table~\ref{tab:ratios}.
    }
    \label{fig:comparison_profiles}
\end{figure}

A key aspect to consider in the modelling of Ba lines is the determination of the microturbulence ($v_{{mic}}$). 
This parameter has a substantial impact on the determination of the barium abundance in RGB stars. 
More importantly, it affects the determination of isotopic ratios both directly and indirectly—directly by influencing the line profile shapes, and indirectly through its impact on the derived barium abundance.
Therefore, a dedicated and accurate determination of $v_{mic}$—specifically tailored to the Ba lines—is essential.
In resonance lines, lower $v_{mic}$ induce isotopic ratios diagnoses closer to the r-process. 
This is because, for a fixed $v_{mic}$ value,  spectral synthesis requires a source of line enhancement to fit the observational line profile. This could be either the barium abundance or the isotopic ratios. Thus, when a reasonably accurate abundance is determined from subordinate lines (which are insensitive to isotopic ratios variation), and it is fixed to determine the isotopic ratios in resonance lines, the only source of line enhancement is the isotopic ratios.
Therefore, since the ratios corresponding to the r-process enhance the line profiles more than those of the s-process (see Fig.~\ref{fig:comparison_profiles}), the isotope diagnoses tend to be biased toward the r-process.
In other words, adopting a low $v_{mic}$ that is not appropriate for resonance lines can lead to misleading diagnoses, artificially biasing the inferred isotopic ratios toward an r-process signature.

In spectral analyses using 1D model atmospheres, $v_{mic}$ is conceived as a velocity field that acts on the line opacity inducing a broadening effect that most strongly affects lines of moderate strength \citep[e.g.][]{takeda2022SoPh..297....4T}, i.e. those in the flat part of the curve of growth (see Fig.~\ref{fig:growth}).
Because of this, $v_{mic}$ is often
calibrated by requiring the abundances obtained from weak and moderately intense lines to be consistent regardless of their EW or REW.
\cite{steffen2013MSAIS..24...37S} provided an LTE $v_{mic}$ validation for RGB stars of solar metallicity based on 3D atmosphere models, where
values correlated with \teff\ remaining between 1 and 1.4~km~s$^{-1}$ are obtained.
\cite{dutra-ferreira2016A&A...585A..75D}
have done a similar work offering an empirical relation as function of \teff\ and \logg\ based on 3D models —for stars with solar metallicity—, which returns no larger values than 1.75~km~s$^{-1}$ for stars with \logg\ $< 1$~dex. 
However, works in the literature often show values close or slightly larger than 2~ km~s$^{-1}$ \citep[e.g.][among many others]{hansen2015ApJ...807..173H,hansen2018ApJ...858...92H,aoki2025PASJ..tmp...22A}.
For instance, in the analysis of the best studied metal-poor RGB star, the Gaia Benchmark HD~122563, four spectroscopic algorithms provide consistent values of $v_{mic}$ derived from iron lines close to 1.9~km~s$^{-1}$ \citep{jofre2014A&A...564A.133J}.

Even assuming the ideal scenario in which $v_{mic}$ is accurately determined from Fe lines, the crucial question is whether this value is also appropriate for modelling Ba lines, given their different line strengths and formation depths.
The likely reason for a negative answer is the influence of chromospheric layers, which can alter the formation of strong Ba lines in ways not captured by standard photospheric models typically used to derive $v_{mic}$ from Fe lines.
A detailed discussion on this topic is provided by \citet{reddy2017ApJ...845..151R}, who examine evidence related to the formation depths of strong Ba lines. They expose that, for the solar case, 
the depths based on calculations using 1D models \citep{mashonkina1999A&A...343..519M, mashonkina2006A&A...456..313M} appear significantly deeper than those derived from empirical photospheric models tailored to the Sun \citep{gurtovenko2015arXiv150500975G}. 
The authors consider this discrepancy as an evidence of  barium abundance biases induced by the simplified hydrostatic approximation in stars with spectra modulated by their chromosphere.
In most RGB stars, the subordinate Ba lines are oversaturated (REW $\gtrsim -4.6$), whereas the resonance lines remain in the damping region of the curve of growth (REW $\gtrsim -4.5$; see Fig.~\ref{fig:growth}).
These lines are notably strong, and their cores therefore probe higher atmospheric layers than those of saturated Fe lines. As a result, the influence of chromospheric layers on the barium lines may manifest through an irregular response to the microturbulent velocity derived from Fe lines.
This problem has been known as {\it the barium puzzle} \citep{dorazi2009ApJ...693L..31D,dorazi2012MNRAS.423.2789D}, manifested as barium abundance excess observed especially in RGB stars and young dwarfs relative to old dwarfs in open clusters of solar-like metallicity \citep[e.g.][]{ Baratella2020A&A...634A..34B,Baratella2021A&A...653A..67B}.
The analysis of \citet{reddy2017ApJ...845..151R} shows 
a correlation between the Ba excess and the chromospheric activity of young stars, but it can be present also in non-active stars as an effect of the incorrect choice of the microturbulent velocity in the spectral analysis.  
Therefore, the apparent Ba excess measurements are probably biases related to spectroscopic methods.

In this work, our goal is to establish a well-calibrated method for determining microturbulence for the different Ba lines of a wide range of strength, and to provide a straightforward prescription to infer it from the microturbulence typically obtained from Fe lines.
We confirm, with observational evidence, that strong Ba lines require $v_{mic}$ values different to those from Fe lines, as previously indicated by \cite{reddy2017ApJ...845..151R}.

In this first paper (Paper~I), we present the analysis of a sample of member stars in the globular cluster NGC~6752, taking advantage of their expected homogeneity in age and metallicity. The sample includes both dwarf and giant stars, allowing us to calibrate the microturbulence parameter required to obtain consistent Ba abundances along the cluster’s evolutionary sequence.
\textcolor{black}{As shown  in Fig.~\ref{fig:growth}, the strength of resonance line we analyse ($\lambda$4934~\AA) in our stars cover the three regions of the curve of growth. The damping region is the one of greatest interest, since most metal-poor stars (including CEMP ones) are expected to have resonance lines of similar strength.}

In Paper~II, we will apply the calibrated relations between stellar parameters and microturbulence to derive Ba abundances and isotopic ratios in a larger sample of metal-poor field stars. Finally, in Paper~III, we will analyse a sample of 14 globular clusters, as presented in \cite{Schiappacasse-Ulloa2025}, deriving Ba abundances and isotopic ratios, and correlating them with the clusters' stellar populations.

The structure of the paper is as follows:
In Sect.\ref{sec:data}, we describe the selection of our stellar sample. Sect.\ref{sec:parameters} outlines the derivation of the stars' atmospheric parameters. In Sect.\ref{sec:adapted_micro}, we detail the method used to calibrate the microturbulent velocity specifically for barium lines. Sect.\ref{sec:Ba_determination} presents the derived barium abundances and isotopic ratios, and explores how these diagnostics may relate to the hypothesis of multiple stellar populations. In Sect.\ref{sec:diffusion}, we compare our metallicity scale with values reported in the literature and discuss the implications in the context of atomic diffusion. Finally, Sect.\ref{sec:conclusions} summarises our main findings and conclusions.

\section{Data and sample}
\label{sec:data}

\begin{figure}
    \centering
    \includegraphics[width=0.8\linewidth]{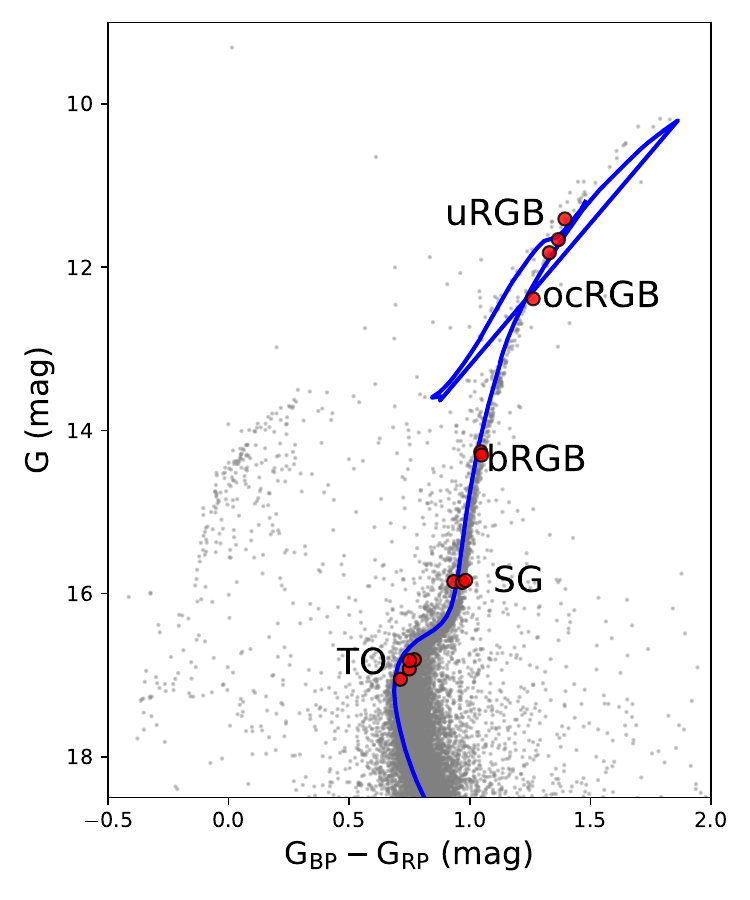}
    \caption{\tiny Colour-magnitude diagram of NGC~6752. Grey and red symbols represent stars in the cluster field and the sample analysed in the present work, respectively. The photometry for both samples was taken from {\it Gaia} e{\sc dr3}.
    An isochrone of 12~Gyr is overplotted as a reference.
    }
    \label{fig:cmd}
\end{figure}

We present a sample of 14 stars belonging to the globular cluster NGC~6752, spanning a broad range of evolutionary stages, from dwarf main sequence turn-off to RGB stars. 
Specifically, we selected stars in four regions of the Kiel diagram: the turn-off (TO), the subgiant (SG) branch, the base of the RGB (bRGB), over the clump RGB (ocRGB), and upper RGB (uRGB); the locations of these regions are indicated in Fig.~\ref{fig:cmd}. The grey dots represent stars located in the field of NGC~6752, selected from the Gaia e{\sc dr3} data within a circular radius of 10 arcmin. We note that these stars were intended to guide the eye and are not necessarily members of the cluster. Red dots represent the stars analysed in the present article, and the blue line is the PARSEC isochrone \citep{Bressan2012} for [M/H]=-1.25 dex, [$\alpha$/Fe]=0.35 dex, an age of 12.0 gigayears (Gyr), (m-M)$_0$=13.13, and an extinction of 0.04.
The whole sample has Ultraviolet and Visual Echelle Spectrograph (UVES) spectra retrieved from the ESO archival data, covering a wavelength range of 480–680 nm. With the exception of 19111828-6000139, these stars were analysed by the {\it Gaia}-ESO survey \citep{Gilmore2022,Randich2022} as part of its calibration program, which also included several open clusters, benchmark stars, and asteroseismic targets. For a detailed overview of the calibration sample, as well as the global calibration and survey homogenisation process, we refer the reader to \citet{Pancino2017A&A...598A...5P} and \citet{Hourihane2023A&A...676A.129H}. Among the stars in this cluster observed by {\it Gaia}-ESO, we selected those with spectra having the highest signal-to-noise ratio (S/N). 
To validate the chemical abundances determined for our two TO stars, we also analysed spectra from two additional TO stars, despite their lower S/N ratios.

To ensure that the selected stars are bona fide members of NGC~6752, we included only those with a membership probability (M-PROB) greater than 0.99, following  \citet{Vasiliev2021}. These findings align with the membership probabilities reported in {\it Gaia}-ESO Data Release 5.1 ({\sc dr5.1}) \citep{Hourihane2023A&A...676A.129H}, which were determined by combining proper motions and parallaxes from {\it Gaia} with radial velocities derived from {\it Gaia}-ESO. For further details on the membership determination process, we refer to \citet{Jackson_2022}.

Table \ref{tab:BasicInfo} provides key information on our sample stars, including coordinates (RA and DEC), [Fe/H], and radial velocity (V$_{\rm r}$) from {\it Gaia}-ESO, proper motions and parallaxes ($\overline{\omega}$) from {\it Gaia} e{\sc dr3}, and S/N.

\begin{table*}
\tiny
\caption{NGC~6752 sample information.}
    \centering
    \begin{tabular}{ccccccccccccc}
    \hline
    \hline
        ID      & Nickname         &   RA    &   DEC   & $\overline{\omega}$& pmra  & pmdec & G & G$_{BP}$-G$_{RP}$ & V$_{\rm r}$ & M-PROB & S/N & POP \\ \hline
        19104537-5958012 & Pavo uRGB-1  & 287.688 & -59.966 & 0.18 & -4.05 & -4.05 & 11.66 & 1.37  & -27.52 & 0.99 & 226 & FG \\
        19104883-5959046 & Pavo uRGB-2  & 287.703 & -59.984 & 0.28 & -3.89 & -4.43 & 11.41 & 1.39  & -30.24 & 0.99 & 186 & FG \\ 
        19111945-6000347 & Pavo ocRGB-1 & 287.831 & -60.009 & 0.26 & -3.63 & -3.89 & 11.82 & 1.33  & -30.30 & 0.99 & 157 & SeG \\
        19104475-5952179 & Pavo ocRGB-2 & 287.686 & -59.871 & 0.25 & -3.10 & -4.08 & 12.39 & 1.26  & -27.14 & 0.99 & 120 & FG \\ 
        19110566-5957041 & Pavo bRGB-1  & 287.773 & -59.951 & 0.23 & -3.52 & -3.65 & 14.26 & 1.05  &    --- & 0.99 & 86  & SeG \\ 
        19104473-6000537 & Pavo bRGB-2  & 287.686 & -60.014 & 0.20 & -3.19 & -3.93 & 14.30 & 1.05  &    --- & 0.99 & 96  & FG \\ 
        19103672-6002011 & Pavo SG-1    & 287.652 & -60.033 & 0.21 & -3.07 & -4.10 & 15.85 & 0.93  &    --- & 0.99 & 76  & SeG \\ 
        19110501-5955274 & Pavo SG-2    & 287.770 & -59.924 & 0.14 & -3.19 & -3.98 & 15.86 & 0.97  &    --- & 0.99 & 98  & --- \\ 
        19111828-6000139 & Pavo SG-3    & 287.826 & -60.003 & 0.33 & -3.09 & -4.31 & 15.84 & 0.98  &    --- & 0.99 & 67  & --- \\
        19102025-5958306 & Pavo SG-4    & 287.584 & -59.975 & 0.04 & -3.12 & -3.85 & 15.77 & 15.77 &    --- & 0.99 & 110 & FG \\
        19102677-6003089 & Pavo TO-1    & 287.611 & -60.052 & 0.16 & -2.89 & -3.64 & 16.92 & 0.75  &    --- & 0.99 & 71  & SeG \\ 
        19105986-6002171 & Pavo TO-2    & 287.749 & -60.038 & 0.17 & -2.88 & -3.59 & 17.05 & 0.71  &    --- & 0.99 & 71  & FG \\ 
        19112283-6002062 & Pavo TO-3    & 287.845 & -60.035 & 0.18 & -3.12 & -4.17 & 16.81 & 0.77  & -30.46 & 0.99 & 28  & FG \\
        19115079-5957477 & Pavo TO-4    & 287.961 & -59.963 & 0.20 & -2.71 & -4.29 & 16.82 & 0.75  & -28.95 & 0.99 & 27  & SeG\\
        \hline
    \end{tabular}
    \begin{tablenotes}
    \item{} \textbf{Notes.} {Position (RA and DEC), $\overline{\omega}$, the astrometric information (proper motions and V$_{\rm r}$) and photometry collected from {\it Gaia} e{\sc dr3} \citep{Gaia_dr3}. The column M-PROB lists the membership probability reported by \citet{Vasiliev2021} for each selected star.
    S/N lists the spectral signal-to-noise ratio, and the column POP lists the population generation (related discussion in Sect. ~\ref{sec:MSTO}).
    } 
    \end{tablenotes}
    \label{tab:BasicInfo}
\end{table*}

\section{Atmospheric parameters}
\label{sec:parameters}
As a first step, we redetermined the stellar parameters of our sample using techniques that deliver high precision for every parameter.
\begin{table*}
\tiny
\caption{Atmospheric parameters of the NGC~6752 stars.}
    \centering
    \begin{tabular}{lcccccccccc}
    \hline
    \hline
        Nickname & \teff\  &   \logg$^{LTE}$  &  \logg$^{NLTE}$ & [Fe/H]$^{fit,LTE}$ &  [Fe/H]$_{LTE}$   & [Fe/H]$_{NLTE}$ & $v_{\text{mic}}^{fit,LTE}$ & $v_{mic}^{LTE}$  & $v_{mic}^{NLTE}$    \\
        \hline
        Pavo uRGB-1 & $4509 \pm 30$ & $1.15 \pm 0.15$ & $1.50 \pm 0.15$ & $-1.50$ & $-1.53 \pm 0.10$ & $-1.35 \pm 0.12$ & 1.60 & $1.56 \pm 0.07$ & $1.59 \pm 0.10$ \\
        Pavo uRGB-2 & $4479 \pm 30$ & $1.00 \pm 0.15$ & $1.40 \pm 0.15$  & $-1.49$ & $-1.54 \pm 0.11$ & $-1.35 \pm 0.13$ & 1.49 & $1.48 \pm 0.10$ & $1.45 \pm 0.15$ \\ 
        Pavo ocRGB-1 & $4624 \pm 82$ & $1.50 \pm 0.15$ & $1.90 \pm 0.15$ & $-1.43$ & $-1.52 \pm 0.12$ & $-1.32 \pm 0.13$ & $1.52$ & $1.71 \pm 0.10$ & $1.72 \pm 0.14$ \\ 
        Pavo ocRGB-2 & $4691 \pm 50$ & $1.60 \pm 0.15$ & $1.90 \pm 0.15$ & $-1.43$ & $-1.49 \pm 0.11$ & $-1.27 \pm 0.15$ & 1.36 & $1.44 \pm 0.07$ & $1.35 \pm 0.11$ \\ 
        Pavo bRGB-1 & $5153 \pm 77$ & $2.70 \pm 0.15$ & $2.90 \pm 0.15$ & $-1.33$ & $-1.38 \pm 0.12$ & $-1.19 \pm 0.17$ & 1.13 & $1.27 \pm 0.11$ & $1.02 \pm 0.14$ \\ 
        Pavo bRGB-2 & $5197 \pm 30$ & $2.60 \pm 0.15$ & $2.90 \pm 0.15$ & $-1.37$ & $-1.43 \pm 0.14$ & $-1.20 \pm 0.16$ & 1.14 & $1.35 \pm 0.10$ & $0.79 \pm 0.18$  \\ 
        Pavo SG-1 & $5437 \pm 30$ & $3.30 \pm 0.15$ & $3.60 \pm 0.15$ & $-1.31$ & $-1.37 \pm 0.13$ & $-1.23 \pm 0.14$ & 1.10 & $1.36 \pm 0.12$ & $1.02 \pm 0.22$ \\ 
        Pavo SG-2 & $5351 \pm 71$ & $3.20 \pm 0.15$ & $3.40 \pm 0.15$ & $-1.46$ & $-1.46 \pm 0.14$ & $-1.27 \pm 0.17$ & 1.27 & $1.27 \pm 0.13$ & $1.14 \pm 0.16$ \\
        Pavo SG-3 & $5320 \pm 68$ & $3.20 \pm 0.15$ & $3.55 \pm 0.15$ & $-1.50$  & $-1.55 \pm 0.09$  & $-1.43 \pm 0.12$  & 1.36 & $1.53 \pm 0.09$ & $1.42 \pm 0.12$ \\ 
        Pavo SG-4 & $5343 \pm 30$ & $3.25 \pm 0.15$ & $3.55 \pm 0.15$ & $-1.46$ & $-1.46 \pm 0.14$ & $-1.37 \pm 0.16$ & 1.29 &  $1.41 \pm 0.13$ & $1.20 \pm 0.19$ \\
        Pavo TO-1 & $6090 \pm 95$ & $3.60 \pm 0.15$ & $3.90 \pm 0.15$ & $-1.60$ & $-1.60 \pm 0.12$ & $-1.49 \pm 0.13$ & --- & $1.36 \pm 0.22$ & $1.06\pm 0.30$ \\ 
        Pavo TO-2 & $6201 \pm 60$ & $3.70 \pm 0.15$ & $4.00 \pm 0.15$ & $-1.62$ & $-1.67 \pm 0.10$ & $-1.60 \pm 0.11$ & --- & $1.61 \pm 0.23$ & $1.36 \pm 0.32$ \\ 
        Pavo TO-3 & $6145 \pm 359$ & $4.00 \pm 0.15$ & $4.20 \pm 0.15$ & $-1.59$ & $-1.53 \pm 0.15$ & $-1.44 \pm 0.17$ & --- & $1.60 \pm 0.50$ & $1.49 \pm 0.50$\\
        Pavo TO-4 & $6074 \pm 106$ & $4.00 \pm 0.15$ & $3.80 \pm 0.15$ & $-1.65$ & $-1.55 \pm 0.22$ & $-1.42 \pm 0.23$ & --- & $0.80 \pm 0.50$ & $0.60 \pm 0.50$\\
        \hline
    \end{tabular}
    \begin{tablenotes}
    \item{} \textbf{Notes.} {\teff, \logg$^{NLTE}$, and [Fe/H]$_{NLTE}$ are determined to be our accurate parameters. Other quantities derived by diverse assumptions and methods are listed for comparison. The [Fe/H] scale has its zero-point at the solar abundance A(Fe) = 7.45~dex of \cite{grevesse2007SSRv..130..105G}.
    } 
    \end{tablenotes}
    \label{tab:atmo_param}
\end{table*}

\begin{figure}
    \centering
    \includegraphics[width=0.8\linewidth]{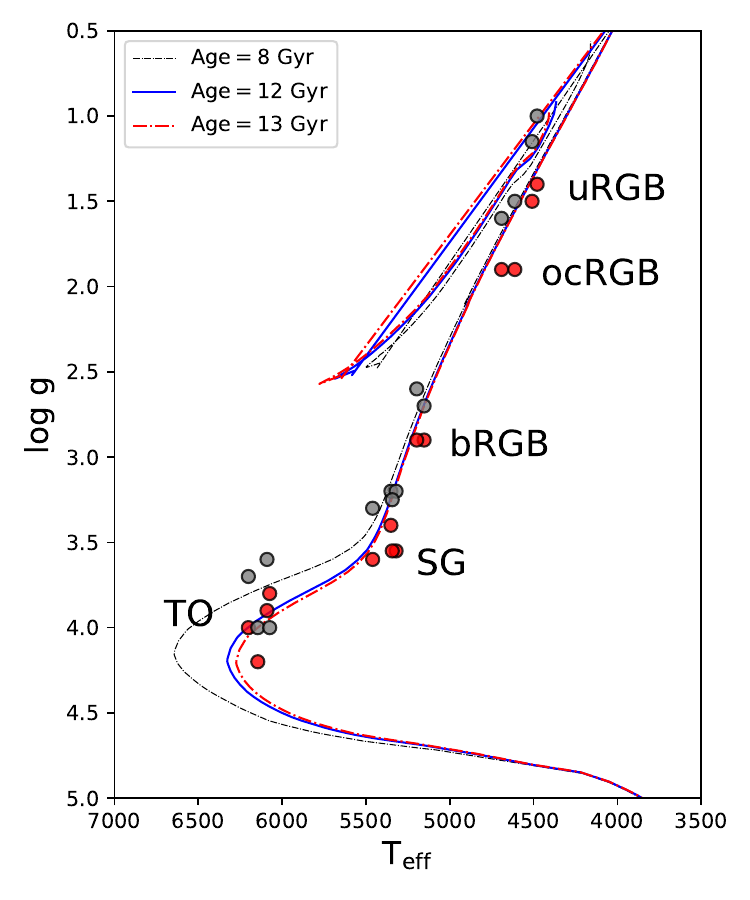}
    \caption{
    \tiny Kiel diagram of the program stars. Red circles display \teff\ and \logg$^{NLTE}$.
    Gray circles display \logg$^{LTE}$.
    Quantities are listed in Table~\ref{tab:atmo_param}.
    }
    \label{fig:kiel}
\end{figure}

\subsection{Determination of \teff}

We determined \teff\ by averaging the outcomes of two 
methods: Balmer H$\alpha$ line modelling and applying the Gaia colour-\teff\ relations of \cite{casagrande2021} based on the InfraRed Flux Method  \citep[IRFM,][]{blackwell1979MNRAS.188..847B}; we refer to the latter as photometric \teff.
We included the latter because the spectra of our TO dwarf stars have low S/N ratios, and thus their H$\alpha$ \teff\ might be less precise.The \teff\ scales of both methods are consistent \citep{giribaldi2021A&A...650A.194G,giribaldi2023A&A...679A.110G}; thus, any differences in the values derived here likely arise from external sources of uncertainty, such as photometric errors, extinction, or residual instrumental patterns in the spectra.

The observational H$\alpha$ profiles were normalized and fitted to 3D NLTE synthetic grids \citep{amarsi2018} using the method of \cite{giribaldi2019A&A...624A..10G,giribaldi2021A&A...650A.194G,giribaldi2023A&A...679A.110G}. 
We derived photometric \teff\ from the Gaia colours  $G_{\rm BP} - G_{\rm RP}$ and $G - G_{\rm BP}$, as these have been shown to provide the most accurate results \citep{giribaldi2023A&A...679A.110G}.
The colours were corrected by subtracting the extinction  $E(B-V) = 0.04$ determined by \cite{gratton2001A&A...369...87G}, which was transformed into the Gaia system using the coefficients of \cite{fitzpatrick1999PASP..111...63F}, as explained in \cite{casagrande2021} and implemented in the COLTE\footnote{\url{https://github.com/casaluca/colte}.} routine.
The final \teff\ is computed as the weighted mean of the temperatures derived by the two methods, with the weights given by the inverse square of their respective uncertainties.
The uncertainties of the temperatures derived from H$\alpha$ were estimated based on the fitting errors associated with the spectral noise. For the uncertainties of the temperatures derived from the colour–\teff\ relations, we adopted a constant error of 100~K, which accounts for the intrinsic precision of the relations themselves (70–80~K) and the combined uncertainties in the photometric colours and extinction estimates ($\sim$30~K).
Total errors are computed following Eq.~1 in \cite{giribaldi20252025A&A...698A..11G} considering a minimum error of  30~K.
The photometric temperatures derived from $G - G_{\rm BP}$ for some stars differ significantly from those obtained via $G_{\rm BP} - G_{\rm RP}$ and H$\alpha$. This discrepancy may result from biases in the measured magnitudes.
Table~\ref{tab:teffs} lists the stellar temperatures derived using each method.

\subsection{Determination of \logg}
We determined \logg\ by searching for the excitation equilibrium of neutral and ionised Fe lines under NLTE (\logg$^{NLTE}$);  we also determined \logg\ under LTE for comparison (\logg$^{LTE}$). 
For that, we performed line synthesis using the radiative transfer code Turbospectrum~2020\footnote{\url{https://github.com/bertrandplez/Turbospectrum_NLTE}}
\citep{gerber2023} with MARCS model atmospheres \citep{gustafson2008}. We considered the atomic parameters from  \cite{heiter2021A&A...645A.106H} and the iron NLTE departure coefficients based on the model atom developed in \cite{bergemann2012} and \cite{semenova2020}.
We assumed excitation equilibrium to compute the surface gravity, following the findings of \cite{giribaldi2023A&A...679A.110G}, who showed that \ion{Fe}{i} and \ion{Fe}{ii} lines yield consistent NLTE abundances when \logg\ is fixed to the value independently derived from the Mg~I~b triplet.

\subsection{Determination of [Fe/H] and $v_{mic}$}
We determined metallicity and $v_{mic}$ applying spectral synthesis to the lines contained in the linelists of \cite{jofre2014A&A...564A.133J}. For giants, we used the linelist adopted for HD~122563 and HD~220009 in that paper, to which we added six \ion{Fe}{ii} lines from \cite{melendez2009A&A...497..611M}, namely the lines at $\lambda$ 5018.43, 5169.028, 5234.62, 5362.86, 5425.26, and 5534.83~\AA.
For dwarfs, we combined the linelists used for metal-poor stars in \cite{jofre2014A&A...564A.133J}, to which we added the six lines above.
Figure~\ref{fig:kiel} shows the location of our sample stars in the Kiel diagram, in which the evolutionary stages are labelled. 
Isochrones with slightly varying ages and metallicities are overplotted for reference. The diagram compares \logg$^{NLTE}$ and \logg$^{LTE}$, illustrating that assuming excitation equilibrium under LTE results in significantly underestimated surface gravities. Our final adopted stellar parameters are \teff, \logg$^{NLTE}$, and [Fe/H]$_{NLTE}$.

\subsection{Evaluation of biases in the determination of $v_{mic}$}
Our ultimate goal is to assess whether $v_{\text{mic}}$ derived from Fe lines can be reliably applied to both subordinate and resonance Ba lines, or whether it requires adjustment through calibrated relations—under the initial assumption that the stars in the cluster exhibit a negligible intrinsic spread in barium abundance.

The first step is to evaluate potential biases and to identify and exclude outlier lines,  aiming to define the best set of lines to derive $v_{mic}$. To this end, we derived [Fe/H] and $v_{\text{mic}}$ from the iron lines using two different methods.
The former method is based on the application of the ionization balance. It consists of deriving an average [Fe/H] from individual line-by-line measurements. In this process, the \teff\ is kept fixed, while the \logg, $v_{mic}$, and macroturbulent velocity ($v_{mac}$) are allowed to freely vary. These parameters are adjusted until no correlation is found between the individual [Fe/H] values and REWs.
We imposed an upper limit to REW < $-4.8$ to exclude oversaturated lines. 
We provided both LTE ($v_{mic}^{LTE}$) and NLTE ($v_{mic}^{NLTE}$) determinations from this method.
In the latter method, we performed a global fit by simultaneously fitting all Fe lines with REW less than $-4.8$, using LTE synthetic spectra. In this approach, [Fe/H]$^{fit,LTE}$,  $v_{mic}^{fit,LTE}$, and  $v_{mac}$ were treated as free parameters, while \logg$^{\mathrm{LTE}}$ and \teff\ were kept fixed.
All parameters are listed in Table~\ref{tab:atmo_param}.

We took special care in the spectral normalization because it is one of the most important factors influencing the microturbulence determination. We applied local normalization by selecting wavelength chunks of $\pm$1~\AA\ around the line centre. 
This practice is reliable in spectra of moderate-resolution of metal-poor stars  because their metal lines are frequently surrounded by continuum regions of reasonable extension.
We rejected lines that are too noisy or likely blended with other features. For that, we fitted the lines with Gaussian functions and supervised their fitting quality using the minimum $\chi^2$ test.
We observe for every star, that the dispersion of $\chi^2$ decreases with the wavelength; likely because the S/N increases. 
For giant and subgiant stars, we excluded lines with $\chi^2$ values exceeding the 75th percentile. An example is shown in Fig.~\ref{fig:clipping_lines} in the Appendix.
In spectra of dwarf stars, Fe lines are scarce, not only because their higher temperatures, but also because the spectral noise is higher. 
For these stars, we supervised the fits manually on a line-by-line basis and excluded those with insufficient quality. Thus, we maximised the selection of well-shaped lines obtaining a total of 45 ones. 
Once a spectrum with locally normalised Fe lines is produced and low-quality lines are clipped from the lists, we applied the two methods above to derive [Fe/H] and $v_{mic}$. This approach ensures that any differences in the results are solely attributed to the methods themselves.

\section{Microturbulence adapted to barium lines}
\label{sec:adapted_micro}

The fundamental premise underlying the present calibration work is the theoretical expectation of homogeneity in the Ba abundance of the cluster, irrespective of the evolutionary stage of the stars. 
In reality, the situation proves to be considerably more complex than initially anticipated.
Namely, A(Ba) of TO stars has been reported to be lower than in RGB stars by $\sim$0.15~dex in the metal-poor cluster NGC~6121 \citep{nordlander2024MNRAS.52712120N}.
The microturbulence calibrations we present in this work are computed considering an overall abundance uncertainty that cover such potential variation in case it is authentic.

Additionally, 
the two turn-off stars with the highest S/N (TO-1 and TO-2), initially planed to serve as proxies, exhibit differing barium abundances and isotopic ratios. 
For this reason, in a posterior validation step, we included two additional TO stars, even though their spectra have much lower S/N. These stars were used to identify the anomalous star among TO-1 and TO-2.
In Section~\ref{sec:MSTO}, we briefly discuss some hypotheses regarding its origin, and refer the reader to Paper~III (Schiappacasse Ulloa el al. in prep.) for an in-depth discussion on the relationship between different stellar populations in globular clusters and the isotopic abundance of Ba.

\subsection{Subordinate lines}
\label{sec:calib_subordinate}

\begin{figure*}
    \centering
    \includegraphics[width=0.33\linewidth]{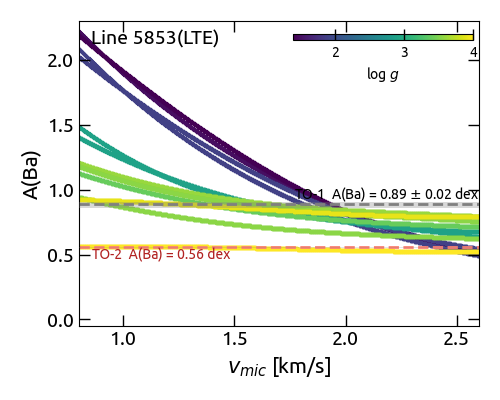}
    \includegraphics[width=0.33\linewidth]{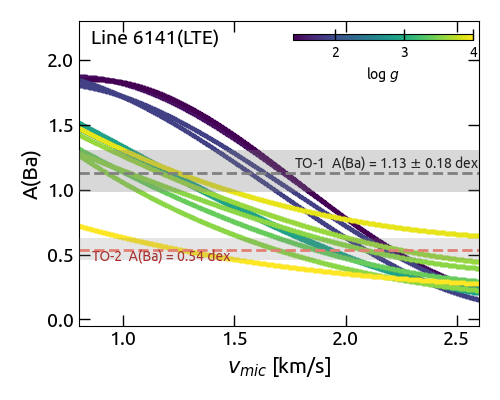}
    \includegraphics[width=0.33\linewidth]{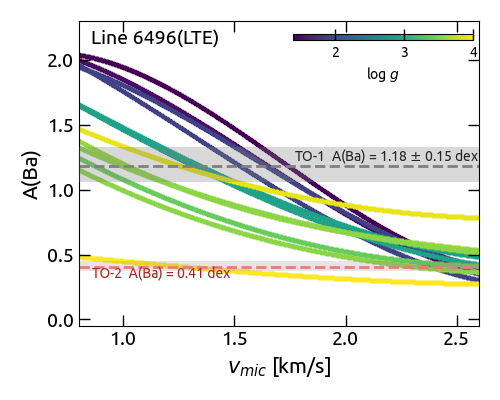}
    \caption{\tiny LTE barium abundance as function of $v_{mic}$ for the subordinate lines 5853, 6141, and 6496~\AA. Each trend is related to one star and is colour-coded according to \logg.
    Shades enclose the most probable ranges of A(Ba) of the TO-1 and TO-2 stars; see main text. Dashed gray and red lines indicate the mean A(Ba) in the shades for each star, respectively.}
    \label{fig:vmic_var}
\end{figure*}

In this section, we investigate the behaviour of the Ba subordinate lines, which are those originating from transitions between excited energy levels, rather than from the ground state.
We determined  A(Ba)\footnote{$A(\mathrm{Ba}) = 12 + \log \left( \frac{N(\mathrm{Ba})}{N(\mathrm{H})} \right)$, where $N(\mathrm{Ba})$ and $N(\mathrm{H})$ are the number densities (abundances by number) of barium and hydrogen, respectively.} in our sample stars by varying $v_{\text{mic}}$ in the range 0.8–2.6 km~s$^{-1}$, in order to quantify the sensitivity of each line to $v_{\text{mic}}$ and to examine how this sensitivity varies with the evolutionary stage, represented here by \logg\ across the four  stages.
Figure~\ref{fig:vmic_var} shows the sensitivity of each subordinate line to variations in $v_{\text{mic}}$; corresponding NLTE results are shown in Fig.~\ref{fig:vmic_var_NLTE} in the Appendix.
The most evident—and expected—result is that, for all lines, the abundance is less sensitive to $v_{\text{mic}}$ in TO stars than in RGB stars.
Considering that in TO stars the Ba lines are weak, thus less affected by departures from 1D hydrostatic atmosphere modelling, we assume that their abundances are the closest to the true values; though in Fig.~\ref{fig:vmic_var_NLTE} we show A(Ba) under 1D NLTE is about $-0.2$~dex relative to that in 1D LTE.
Therefore, the barium abundances of TO-1 and TO-2 might be adopted as the best proxies for the cluster abundance. The detailed determination of their Ba abundances is presented in Section~\ref{sec:Ba_determination}. However, as anticipated, we noted that the barium abundance of the TO-2 star is significantly lower than that of TO-1. 
Moreover, the abundances of TO‑3 and TO‑4 closely match that of TO‑1, although the lower S/N of their spectra results in lower precision.
This has important implications, as reproducing the same A(Ba) as in TO-2 in the other stars would require unrealistically high values of $v_{\text{mic}}$.

The plots in Fig.~\ref{fig:vmic_var} include error shades associated to the A(Ba) ranges of the TO stars obtained when $v_{mic}$ is changed from 1 to 1.5~km~s$^{-1}$. 
This range was determined 
with the \titan~I metal-poor dwarfs \citep{giribaldi2021A&A...650A.194G}, performing LTE synthesis fixing their quoted parameters. Figure~\ref{fig:vmic_titans} shows the distribution of $v_{mic}$ as functions of the atmospheric parameters. 
Assuming that the shades in Fig.~\ref{fig:vmic_var} enclose the true abundances, we compute for every curve in the plots, the $v_{mic}$ ranges corresponding to the enclosed A(Ba) ranges.

\begin{figure*}
    \centering
    \includegraphics[width=0.33\linewidth]{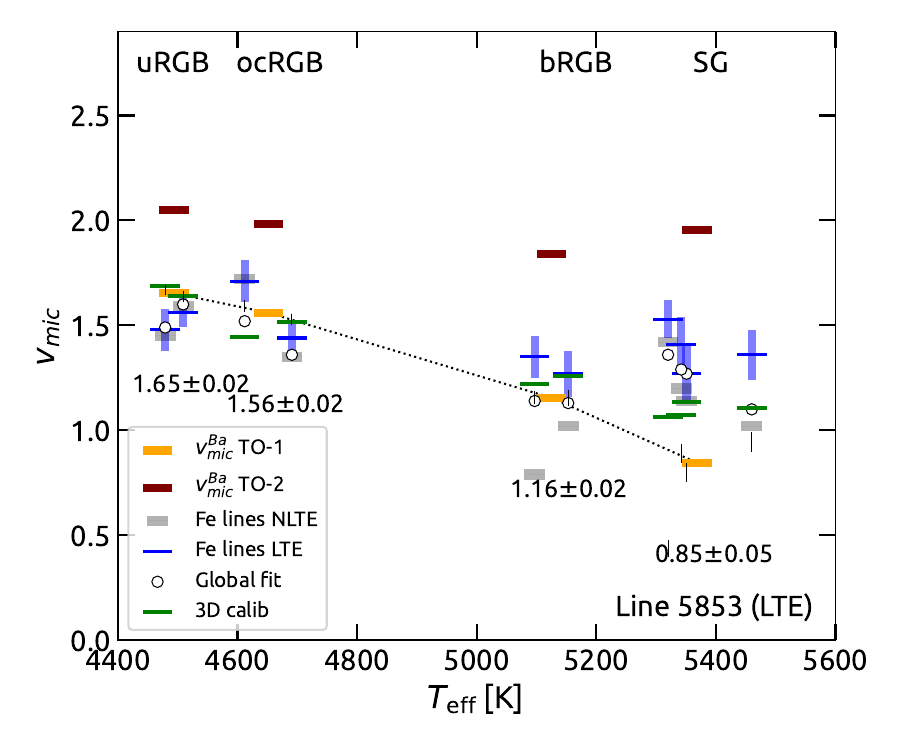}
    \includegraphics[width=0.33\linewidth]{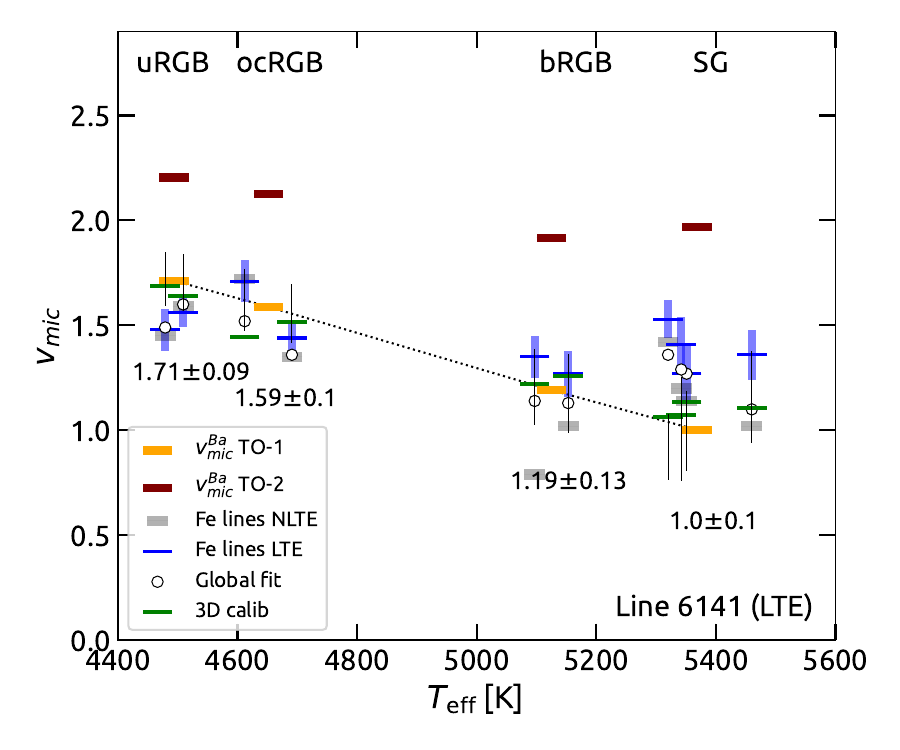}
    \includegraphics[width=0.33\linewidth]{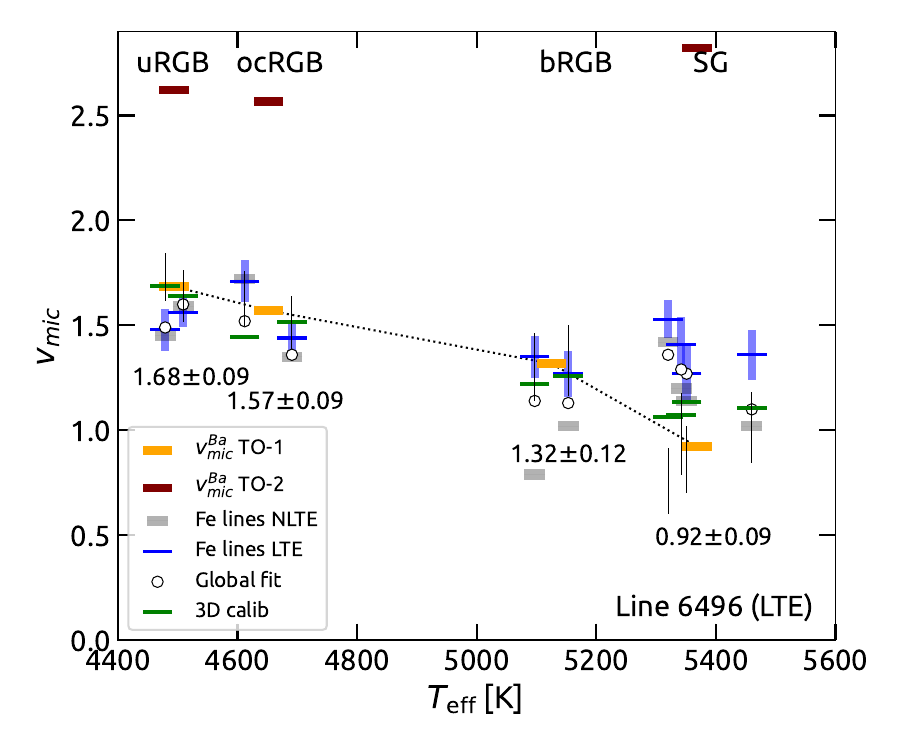}
    \caption{\tiny Microturbulence as function of \teff\ for the stars in Table~\ref{tab:BasicInfo}. 
    Panels from the left to the right show quantities derived from the subordinate lines at $\lambda$5853, 6141, and 6496~\AA, respectively.
    TO stars are excluded on purpose. 
    Vertical bars represent $v_{mic}$ ranges computed by the intersections of the trends with the shades in Fig.~\ref{fig:vmic_var}. 
    Orange dashes are medians of contiguous bars, which correspond to stars in the stages indicated on the top; these values are defined as \vcalibsub.
    Brown dashes represent \vcalibsub\ if it were derived using A(Ba) of the TO-2 star (red dashed line in Fig.~\ref{fig:vmic_var}).
    Blue dashes, gray dashes, and circles represent $v_{mic}^{LTE}$, $v_{mic}^{NLTE}$, and $v_{mic}^{fit}$, respectively.
    Blue bars are the errors of $v_{mic}^{LTE}$. Green dashes are computed with the relation based on 3D models of \cite{dutra-ferreira2016A&A...585A..75D}.
    }
    \label{fig:vmic_teff}
\end{figure*}

Figure~\ref{fig:vmic_teff} shows the $v_{mic}$ ranges computed as functions of \teff\ (black solid bars) when the A(Ba) of the TO-1 star is assumed as proxy. Each column of the plot corresponds, respectively,  to the subordinate lines  at $\lambda$5853, $\lambda$6141, and $\lambda$6496~\AA\ in LTE.
Orange dashes indicate the medians of the mean values derived from the bars within each evolutionary stage. These median values are shown in the plots alongside their associated uncertainties, computed by summing in quadrature the standard deviation of the bar means and the individual bar errors divided by the number of bars. 
These values, hence, represent the microturbulence calibrated to the subordinate lines, hereafter referred to as \vcalibsub. 

Microturbulence from Fe lines listed in Table~\ref{tab:atmo_param} are shown for comparison; only the errors of $v_{mic}^{LTE}$ are plotted.
$v_{\text{mic}}^{fit,LTE}$, $v_{mic}^{LTE}$, and $v_{mic}^{NLTE}$ are equivalent for stars at the uRGB and ocRGB stages. 
At the bRGB and SG, $v_{mic}^{fit,LTE}$
is slightly lower than $v_{mic}^{LTE}$ by 0.1-0.2~km~s$^{-1}$, whereas $v_{mic}^{NLTE}$ tend to be even lower. 
We also over-plot microtubulence values computed by the relation of \cite{dutra-ferreira2016A&A...585A..75D} based on 3D LTE models (green dashes).
An excellent agreement is observed between these values and \vcalibsub\ for all evolutionary stages except for the SG stars.
Therefore, for practical purposes, we recommend the use of the relation in \cite{dutra-ferreira2016A&A...585A..75D}, which is transcribed below in Eq.~\ref{eq:vmic3D}, for safely determining Ba abundances from the subordinate lines:
\begin{equation}
 \begin{aligned}
    v_{mic}^{\mathrm{Ba}} = & \; 0.998 + 3.16 \times 10^{-4} \, X - 0.253 \, Y \\ 
    & \; -2.86 \times 10^{-4} \, X \, Y + 1.65 \, Y^2
 \end{aligned}
 \label{eq:vmic3D}
\end{equation}

\noindent where $X =$~\teff $- 5500$ and $Y =$~\logg\ $-4.0$. The root mean square (rms) scatter of this relation is 0.05~km~s$^{-1}$, according to the paper source. Therefore, we adopt this quantity to compute the abundance errors in Sect.~\ref{sec:Ba_determination}.
Fig.~\ref{fig:vmic_teff} also show \vcalibsub\ when the TO-2 star is adopted as the reference for the cluster A(Ba) (brown dashes).
In this case, the \vcalibsub\ values derived from the line at $\lambda$6496~\AA\ are significantly higher than those obtained from the lines at $\lambda$5853 and $\lambda$6141~\AA, exceeding 2.5~km~s$^{-1}$ at all evolutionary stages. Although this exercise is intentionally illustrative—anticipating atypical values of \vcalibsub—those obtained for the SG stars, reaching approximately 2.85~km~s$^{-1}$, are particularly suspicious given that the lines are only moderately saturated (REW $\approx -4.9$). For bRGB stars, the derived \vcalibsub\ becomes unphysically large ($>10$~km~s$^{-1}$); thus, no corresponding value is shown in the right panel of the plot.

Table~\ref{tab:vmic_errors} presents an example of the variations in A(Ba) induced by changes in $v_{\text{mic}}$ across the different evolutionary stages. These values are derived from the behaviour of the line at $\lambda$6141~\AA, as shown in the middle panel of Fig.~\ref{fig:vmic_teff}, assuming the TO-1 star as the reference proxy.

\subsection{Resonance lines}
\label{sec:vmic_resonance}
In the previous section we determined that A(Ba) from subordinate lines of the uRGB, ocRGB, and bRGB are  compatible with that of  TO-1 using the appropriate value of the microturbulent velocity.  
It is therefore expected that their Ba abundances have the same nucleosynthetic origin, e.g. similar isotopic composition.  
In Sect.~\ref{sec:isotopes}, we demonstrate that the barium abundance in star TO-1 is entirely produced by the s-process, whereas in TO-2 it is predominantly of r-process origin.
Anticipating this result, we perform for the resonance lines an exercise similar to that done for subordinate lines in Fig.~\ref{fig:vmic_var}.
However, instead of fixing only A(Ba) to calibrate the microturbulence, we fix both A(Ba) and the isotopic ratios, testing separately the ratios corresponding to the r- and s-process contributions. In this way, we can also assess which of the TO stars reflects the typical composition of the cluster and which one shows a peculiar abundance pattern.

\begin{figure*}
    \centering
    \includegraphics[width=0.33\linewidth]{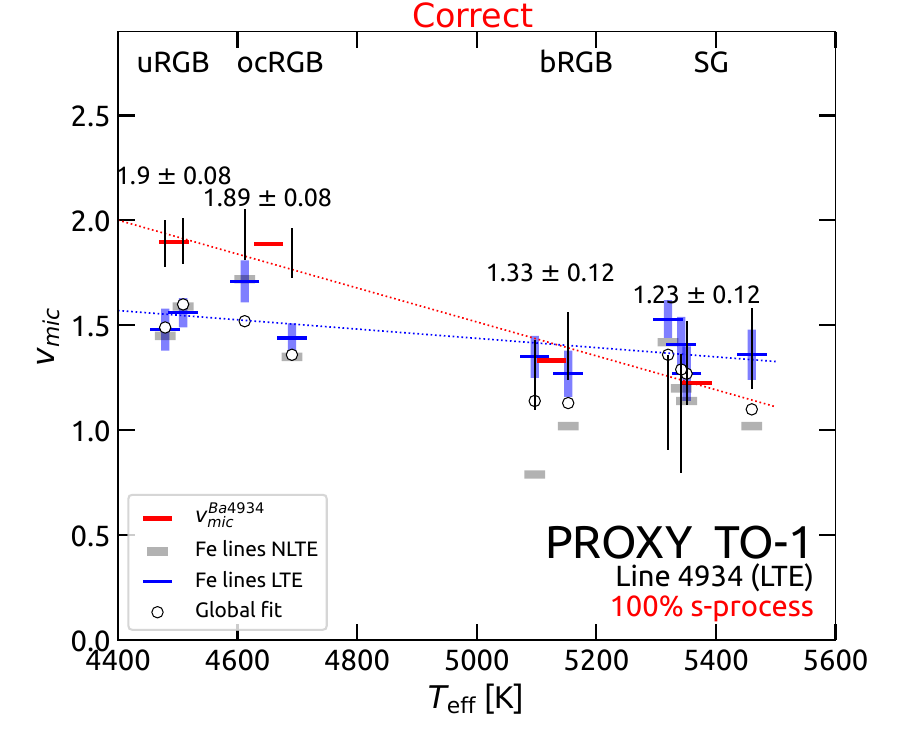}
    \includegraphics[width=0.33\linewidth]{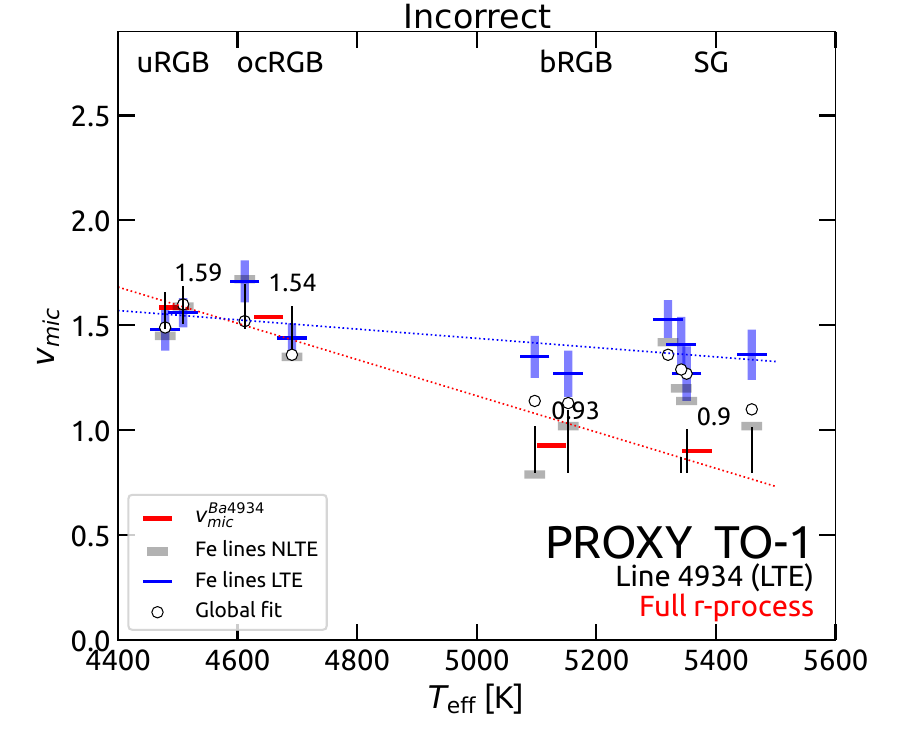}
    \includegraphics[width=0.33\linewidth]{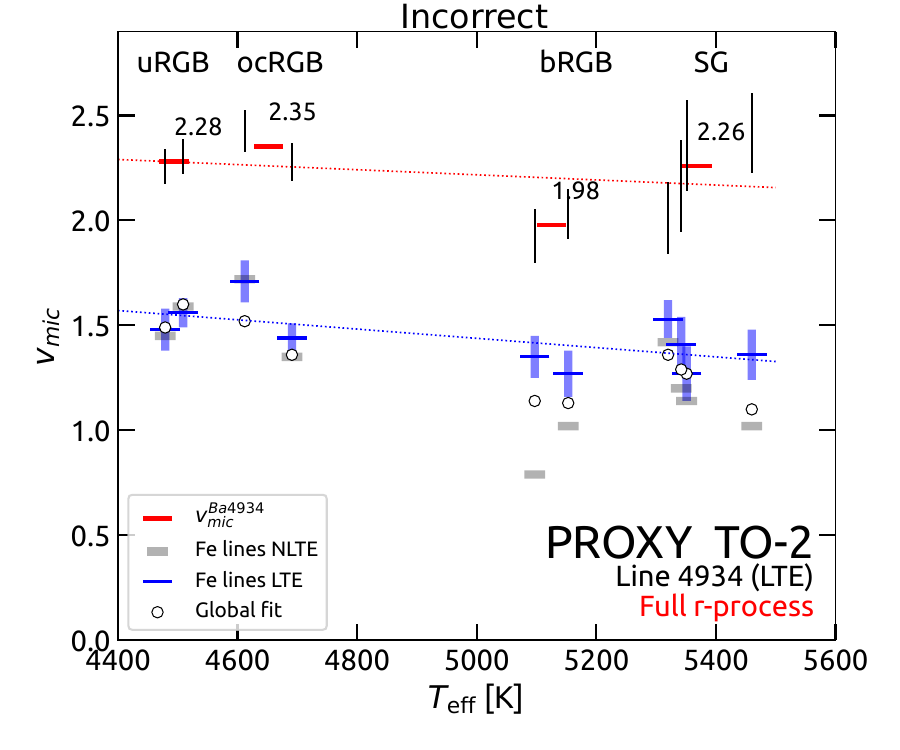}
    \caption{\tiny Microturbulence as function of \teff\ for the line 4934~\AA.
    Left and centre panels shows quantities derived assuming 100 and 0\% s-process  contributions, respectively;     
    the A(Ba) $= 1.15 \pm 0.08$~dex of the TO-1 star is assumed in both cases.
    Right panel shows quantities derived assuming the A(Ba) $= 0.39 \pm 0.08$~dex and full r-process contribution of the TO-2 star.
    Black bars represent the microturbulence ranges required to fit observational lines with synthetic ones.
    Red dashes are the medians of the bars in the stages indicated on the top. 
    The red and blue dotted lines are  linear regressions of the red and blue dashes, respectively.
    }
    \label{fig:vmic_teff_4934}
\end{figure*}

Figure~\ref{fig:vmic_var_4934} in the Appendix displays the trends of A(Ba) as a function of $v_{\rm mic}$, using as reference the A(Ba) values of TO-1 (representative of the s-process) and TO-2 (representative of the r-process). The shaded regions reflect the abundance uncertainties, calculated by adding in quadrature the contributions from $\sigma_{\rm sn}$ and $\sigma_T$. The uncertainty due to $\sigma_v$ was excluded to enhance the precision of our estimates.
We obtain microturbulence adapted to the resonance line at $\lambda$4934~\AA\ (\vcalibres) in an analogous way as in Fig.~\ref{fig:vmic_teff}.
Figure~\ref{fig:vmic_teff_4934} presents a comparison between the microturbulence values derived from Fe lines and those calibrated using Ba lines. The average values of \vcalibres\ and their associated uncertainties —computed following the method described in the previous section— are indicated in each panel.

In the right panel, where the lower A(Ba) value of TO-2 is taken as representative of the cluster abundance, \vcalibres\ exceeds \vlte\ by approximately 0.8~km~s$^{-1}$. Notably, this offset does not follow the expected trend of increasing discrepancy with line strength —from right to left in the plot— as would be predicted due to the greater sensitivity of stronger lines to the limitations of 1D atmospheric models, especially in the damping region of the curve of growth.  More specifically, the central panel, representing the full r-process, shows discrepancies only at higher \teff, where the lines are weaker —contrary to the expected behaviour. Therefore, this calibration attempt is also deemed incorrect.

In contrast, when the higher A(Ba) value of TO-1 is assumed to represent the cluster abundance, the discrepancies minimise. 
The left panel, representing 100\% s-process isotopic ratios, shows that only the strongest lines —from the uRGB and ocRGB stars— require larger microturbulence values than those derived from Fe lines, as expected. Therefore, this calibration can be considered correct.

\begin{figure*}
    \centering
    \includegraphics[width=0.8\linewidth]{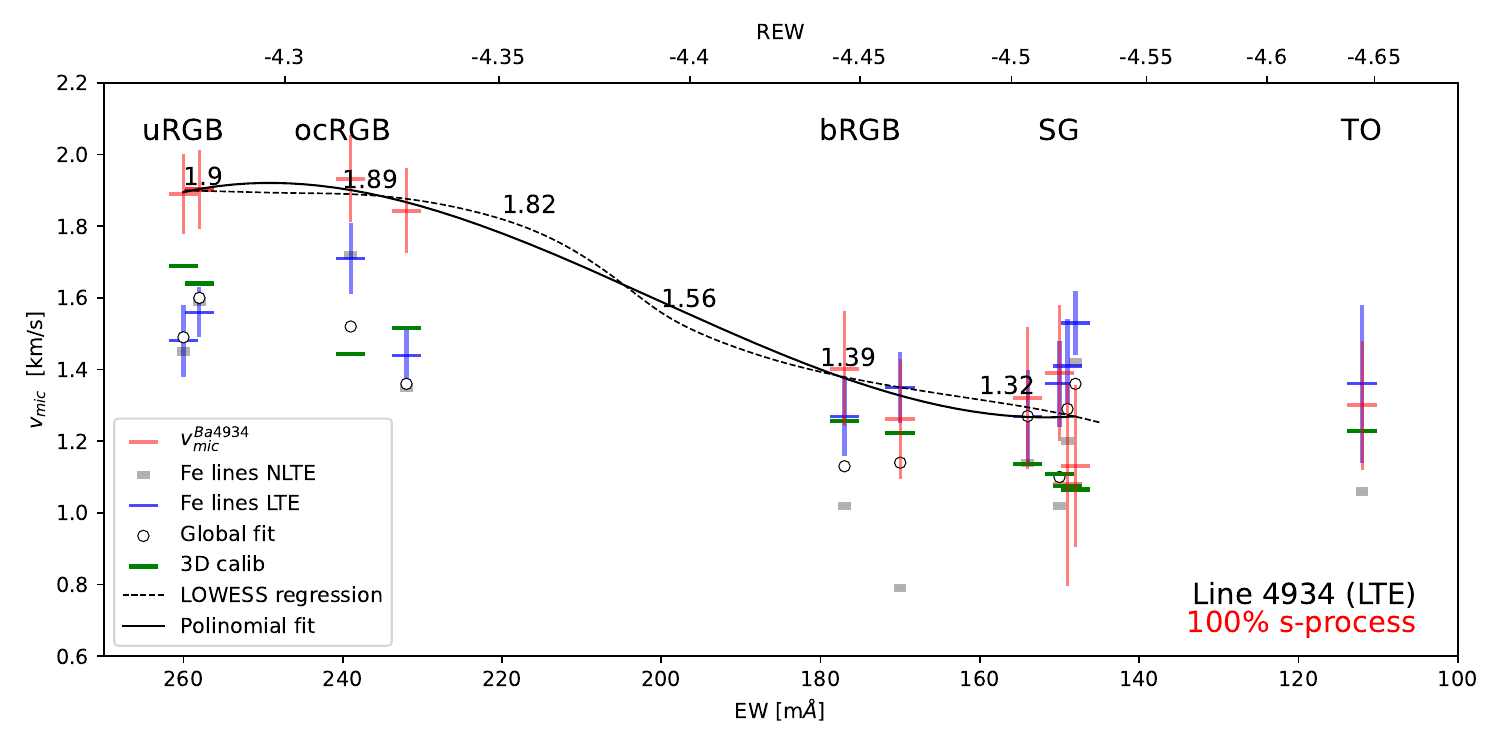}
    
    \caption{\tiny Microturbulence as function of EW and REW for the Ba resonance line at $\lambda$4934~\AA. Microturbulence values determined from diverse methods are represented according to the legends. The solid black line is the polynomial regression of \vcalibres\ (Eq.~\ref{eq:pol}).
    The dashed line is the LOWESS regression. Microturbulence values corresponding to discrete points of the curve, separated by 20 m\AA, from 260 to 160 m\AA\ are noted. 
    }
    \label{fig:vcalib_REW}
\end{figure*}

This supports our choice of considering TO-1 as representative of the cluster's Ba abundance and isotopic ratios, and as the reference for calibrating the microturbulence accordingly. The implications of these findings are discussed in the following section.
A more effective method is to follow the trend shown by the adapted microturbulence as a function of EW and REW in Fig.~\ref{fig:vcalib_REW}. EW and REW are parameters more directly related to line shapes than \teff, making the use of \vcalibres\ less likely to fail in cases of Ba variations.
A LOWESS\footnote{Locally Weighted Scatterplot Smoothing (LOWESS) regressions are applied by the Python {\textit{moepy}} package \cite{LOWESS} available at \url{https://ayrtonb.github.io/Merit-Order-Effect/}, using the parameter frac=0.55.} regression and a polynomial are fitted to \vcalibres\ (red dashes). 
These are represented by the solid and dashed lines; the equation of the latter is given below:

\begin{equation}
 \begin{aligned}
    v_{mic}^{\mathrm{Ba}4934} = & \; 10.935 - 0.15996 \times EW + 0.00084831 \times EW^2 \\ 
    & \; -1.410868 \times 10^{-6} \times EW^3
 \end{aligned}
 \label{eq:pol}
\end{equation}

\noindent 
The uncertainty of the $v_{mic}$ estimates of these trends is of $\pm$0.10~km~s$^{-1}$; it is an approximate quantity to those determined in left panel of Fig.~\ref{fig:vmic_teff_4934}.
Therefore, we adopt this quantity to errors of the isotopic ratios in Sect.~\ref{sec:Ba_determination}.
The trends of the LOWESS and the polynomial function are similar. When these are compared with $v_{mic}$ from 3D models (green dashes, which in well suited for Ba subordinate lines as Fig.~\ref{fig:vmic_teff} shows), it is evident that \vcalibres\ is systematically higher.
At the SG and bRGB stages the difference is about +0.1~km~s$^{-1}$, whereas at the uRGB and ocRGB stages the difference increases to about +0.25~km~s$^{-1}$. 
We recommend to use the polynomial in Eq.~\ref{eq:pol}, or the quantities given in the plot, for fixing $v_{mic}$ when determining Ba isotopic ratios from the line at $\lambda$4934~\AA.

\section{Barium abundance and isotopic ratios}
\label{sec:Ba_determination}

Following the calibration of the microturbulence for the various resonant and subordinate lines of Ba, the abundances of the stars belonging to the cluster can now be determined with a high precision.

\subsection{Barium abundance}
We derived barium abundance assuming LTE conditions. 
We performed spectral synthesis fitting the subordinate lines at $\lambda$5853, 6141, and 6496~\AA.
The atomic parameters of the the two latter can be found in \cite{Gallagher2020A&A...634A..55G}. 
Details on HyperFine Splitting (HFS) and associated oscillator strength (in logarithmic scale, log $gf$) values of these lines are provided in Giribaldi et al. (in prep.).
The abundances of the stars are listed in Table~\ref{tab:ba_abundances}.
The fits are made fixing the microturbulence to the values obtained from Eq.~\ref{eq:vmic3D}. 
Section~\ref{sec:calib_subordinate} provides a justification of this choice.

\begin{table*}
\tiny
\caption{Barium abundances of the NGC~6752 stars.}
    \centering
    \begin{tabular}{lcccccccccc}
    \hline
    \hline
        Nickname &   A(Ba) $\pm \sigma_{sn} \pm \sigma_{T} \pm \sigma_{v} \; (\Sigma_{Ba})$ & s-process  $\pm \sigma_{sn-iso} \pm \sigma_{Ba-iso} \pm \sigma_{T-iso} \pm \sigma_{v-iso} \pm \sigma_{[\mathrm{Fe/H}]} \; (\Sigma_{iso})$ & $v_{mic}^{\mathrm{3D}}$  \\
        & [dex] & [\%] & [km~s$^{-1}$] \\
        \hline
        Pavo uRGB-1 & $1.15 \pm 0.05 \pm 0.02 \pm 0.07 \; (\pm0.09)$ & $100 \pm _{11}^{0} \pm 14 \pm ^{10}_{17} \pm 23 \pm 5 \; (\pm^{28}_{30})$ & $1.64 \pm 0.05$ \\
        Pavo uRGB-2 & $1.15 \pm 0.06 \pm 0.02 \pm 0.07 \; (\pm0.09)$  & $100 \pm ^{0}_{0} \pm 17 \pm ^{10}_{17} \pm 23 \pm 5 \; (\pm^{30}_{33})$ & $1.69 \pm 0.05$  \\ 
        Pavo ocRGB-1 & $1.24 \pm 0.09 \pm 0.07 \pm 0.07 \; (\pm 0.13)$  &  $100 \pm  ^{0}_{0} \pm 26 \pm ^{36}_{63} \pm 23 \pm 4 \; (\pm^{51}_{72})$ & $1.44 \pm 0.05$ \\ 
        Pavo ocRGB-2 & $1.18 \pm 0.08 \pm 0.03 \pm 0.07 \; (\pm0.11)$  &  $100 \pm ^{0}_{6} \pm 23 \pm ^{16}_{28} \pm 23 \pm 5 \; (\pm^{36}_{43}) $  & $1.51 \pm 0.05$ \\
        Pavo bRGB-1 & $1.15 \pm 0.09 \pm 0.05 \pm 0.06 \; (\pm0.12)$ & $71 \pm ^{29}_{71} \pm 26 \pm ^{25}_{42} \pm 12 \pm 5 \; (\pm^{49}_{88})$ & $1.25 \pm 0.05$  \\ 
        Pavo bRGB-2 & $1.11 \pm 0.11 \pm 0.02 \pm 0.06 \; (\pm0.13)$ & $100 \pm ^{0}_{10} \pm 31 \pm ^{10}_{16} \pm 12 \pm 5 \; (\pm^{36}_{39})$ & $1.22 \pm 0.05$   \\ 
        Pavo SG-1 &  $1.19 \pm 0.06 \pm 0.03 \pm 0.04 \; (\pm0.08)$ & $100 \pm _{18}^{0} \pm 17 \pm ^{15}_{25} \pm 12 \pm 6 \; (\pm ^{27}_{32}) $ & $1.11 \pm 0.05$ \\ 
        Pavo SG-2 & $1.09 \pm 0.09 \pm 0.05 \pm 0.04 \; (\pm0.11)$  & $96 \pm ^{14}_{4} \pm 26 \pm ^{21}_{35} \pm 12 \pm 9 \; (\pm ^{32}_{40})$ & $1.14 \pm 0.05$ \\ 
        Pavo SG-3 & $1.00 \pm 0.05 \pm 0.05 \pm  0.04  \; (\pm0.08) $ & $92 \pm ^{28}_{8} \pm 14 \pm ^{20}_{34} \pm 12 \pm 6 \; (\pm^{39}_{39})$ & $1.07 \pm 0.05$ \\ 
        Pavo SG-4 &  $1.13 \pm 0.07 \pm 0.02 \pm 0.04 \; (\pm 0.08)$ & $100 \pm _{2}^{0} \pm 20 \pm ^{9}_{15} \pm 12 \pm 7 \; (\pm^{27}_{30})$ & $1.07 \pm 0.05$ \\
        Pavo TO-1 &  $1.15 \pm 0.03 \pm0.07 \pm 0.03 \; (\pm0.08)$ & $ 94 \pm ^{6}_{22} \pm 9 \pm ^{24}_{32} \pm 12 \pm 7 \; (\pm^{30}_{42})$ & $1.23 \pm 0.05$  \\ 
        Pavo TO-2 &  $0.39 \pm 0.02 \pm 0.05 \pm 0.03 \; (\pm0.06)$ & $0 \pm ^{72}_{0} \pm 6 \pm ^{14}_{19} \pm  12 \pm 7 \; (\pm^{75}_{24})$ & $1.22 \pm 0.05$ \\ 
        Pavo TO-3 & $0.92 \pm 0.12 \pm 0.25 \pm 0.05 \; (\pm 0.28)$ & --- & $1.22 \pm 0.05$\\
        Pavo TO-4 & $1.25 \pm 0.30 \pm 0.08 \pm 0.05 \; (\pm0.31)$ & --- & $1.23 \pm 0.05$ \\
        \hline
    \end{tabular}
    \begin{tablenotes}
    \item{} \textbf{Notes.} {The second column lists the barium abundance in the logarithmic scale $\log N(\rm{H}) = 12$. Abundance errors induced by those of S/N, \teff, and $v_{mic}$ are listed individually (in this order); whereas the total error is given in brackets. Third column lists the isotopic ratios in terms of s-process percentage. Its errors induced by those of the S/N, A(Ba), \teff, $v_{mic}$, and [Fe/H] are listed individually and in this order. The total error is given in brackets. 
    Fourth column lists the microturbulence determined from Eq.~\ref{eq:vmic3D}, which is equivalent to \vcalibsub. 
         } 
\end{tablenotes}
    \label{tab:ba_abundances}
\end{table*}

Total abundance errors are given by the following relation:

\begin{equation}
    \label{eq:error_Ba}
    \Sigma_{Ba} = \sqrt{ \sigma_{sn}^2 + \sigma_T^2 + \sigma_v^2 }
\end{equation}

\noindent
where $\sigma_{sn}$ is the error related to the noise, $\sigma_T$ is the error related to \teff, and $\sigma_v$ is the error related to $v_{mic}$.
All these errors are individually annotated in Table~\ref{tab:ba_abundances}.
Errors induced by typical uncertainties in \logg\ and [Fe/H] are negligible for both dwarfs and giants, i.e lower than 0.01~dex. 
$\sigma_{sn}$ is assumed to be the standard deviation of the A(Ba) values obtained from the subordinate lines. 
$\sigma_T$ is computed as described in  \cite{giribaldi2023A&A...679A.110G} using synthetic spectral grids of the line at $\lambda$6141~\AA\ as a proxy; see associated plots in Fig.~\ref{fig:Ba_dif}.
Errors related to variations of $\pm50$~K are listed in Table~\ref{tab:vmic_errors}.
$\sigma_v$ is estimated from the analysis of Sect.~\ref{sec:adapted_micro}, where the error of the microturbulence ($\pm$0.05~km~s$^{-1}$) is given by the rms of Eq.~\ref{eq:vmic3D}.
We employ this error and the values in Table~\ref{tab:vmic_errors}
to compute $\sigma_v$ for each star.
The values in the table were computed from $v_{mic}$ versus A(Ba) variations of the sensitive line $\lambda$6141~\AA\ (Fig.~\ref{fig:vmic_var}), which was used as proxy.

\begin{table}
\caption{A(Ba) errors induced by \teff\ and $v_{mic}$ errors.}
\label{tab:vmic_errors}
\centering
\tiny 
\begin{threeparttable}
\begin{tabular}{l|ccccccccc}
\hline\hline
Parameter & TO & SG & bRGB & ocRGB & uRGB  \\
\hline
\teff\ & $\pm$0.040 & $\pm$0.035 & $\pm$0.031 & $\pm$0.029 & $\pm$0.029 \\
$v_{mic}$ & $\mp0.033$ & $\mp0.041$ & $\mp0.055$ &  $\mp0.065$ & $\mp0.065$ \\
\hline
\end{tabular}
\begin{tablenotes}
\item{} \textbf{Notes.} {Unities are expressed in dex. Quantities related to \teff\ are computed from the grids represented in Fig.~\ref{fig:Ba_dif} and correspond to a variation of $\pm$50~K.
Quantities related to $v_{mic}$ are extracted from the computations done with the line at 6141~\AA, represented in the middle panel in Fig.~\ref{fig:vmic_var}, and correspond to $v_{mic} \pm 0.05$~km~s$^{-1}$.
} 
\end{tablenotes}
\end{threeparttable}
\end{table}

\subsection{Barium isotopic abundance ratios}
\label{sec:isotopes}

We determined the Ba isotopic ratios by fitting the profile of the resonance line at $\lambda$4934~\AA, fixing the abundances determined from subordinate lines.

This  resonance line is not included in \cite{Gallagher2020A&A...634A..55G}, therefore we computed a $\log{g_if_{ij}} =  g_j \lambda_ {ij}^2 A_{ji}/6.6702\times10^{15} =  -0.172$; where $g_j=2$ is the upper statistical weight of the transition, $\lambda_{ij}=4935.45$~\AA\ is the wavelength of the transition in vacuum, and  $A_{ji}$ is the spontaneous transition probability  from \cite{DeMunshi2015PhRvA..91d0501D}. The detailed isotopic shifts, HFS, and corresponding log $gf$ values for this line are given in Table~\ref{tab:tab1}. They are based on energy levels in the NIST\footnote{\url{https://dx.doi.org/10.18434/T4W30F}} database, and on \cite{1986PhRvA..33.2117S,1993JPhB...26.4289V,2000HyInt.127...57T,2006PhRvA..73b2510I} for hyperfine structure constants of odd \ion{Ba}{ii} isotopes 135 and 137.
We employ the isotopic ratios in Table~\ref{tab:ratios} for the s- and r-processes.

\begin{table*}
\caption{Barium isotopic ratios}
\label{tab:ratios}
\centering
\tiny 
\begin{threeparttable}
\begin{tabular}{l|ccccccccc}
\hline\hline
Process & $^{134}$Ba & $^{135}$Ba & $^{136}$Ba & $^{137}$Ba & $^{138}$Ba \\
\hline
Slow  (s-) & 0.0286 & 0.0222 &  0.0939  & 0.1048  & 0.7505 \\
Rapid (r-) & 0.0000 & 0.3924 & 0.0000 & 0.2690 & 0.3386 \\
Intermediate (i-) & 0.006-0.009 & 0.045-0.027 & 0.041-0.028 & 0.308-0.673 & 0.600-0.263 \\
\hline
\end{tabular}
\begin{tablenotes}
\item{} \textbf{Notes.} {Quantities related to the s- and r-processes are inferred from \cite{goriely2018A&A...609A..29G} and \cite{goriely1999A&A...342..881G}, respectively. 
Isotopic ratios related to the i-process are taken from \cite{martinet2024A&A...684A...8M}. 
} 
\end{tablenotes}
\end{threeparttable}
\end{table*}

Our analysis is based on the premise that heavy elements in the interstellar medium are progressively enriched by nucleosynthetic products. Consequently, we interpret the isotopic ratios as complementary fractions of s- and r-process contributions. We quantify the isotopic ratios in terms of the s-process fraction, ranging from 0\% (indicating a pure r-process origin) to 100\% (a pure s-process origin), as reported in Table~\ref{tab:ba_abundances}. For each star, we assign the isotopic ratio corresponding to the synthetic profile that ({\it i}) reproduces the mean A(Ba) derived from the subordinate lines and ({\it ii}) provides the best fit to the observed line profile.

\begin{figure*}
    \centering
    \includegraphics[width=0.32\linewidth]{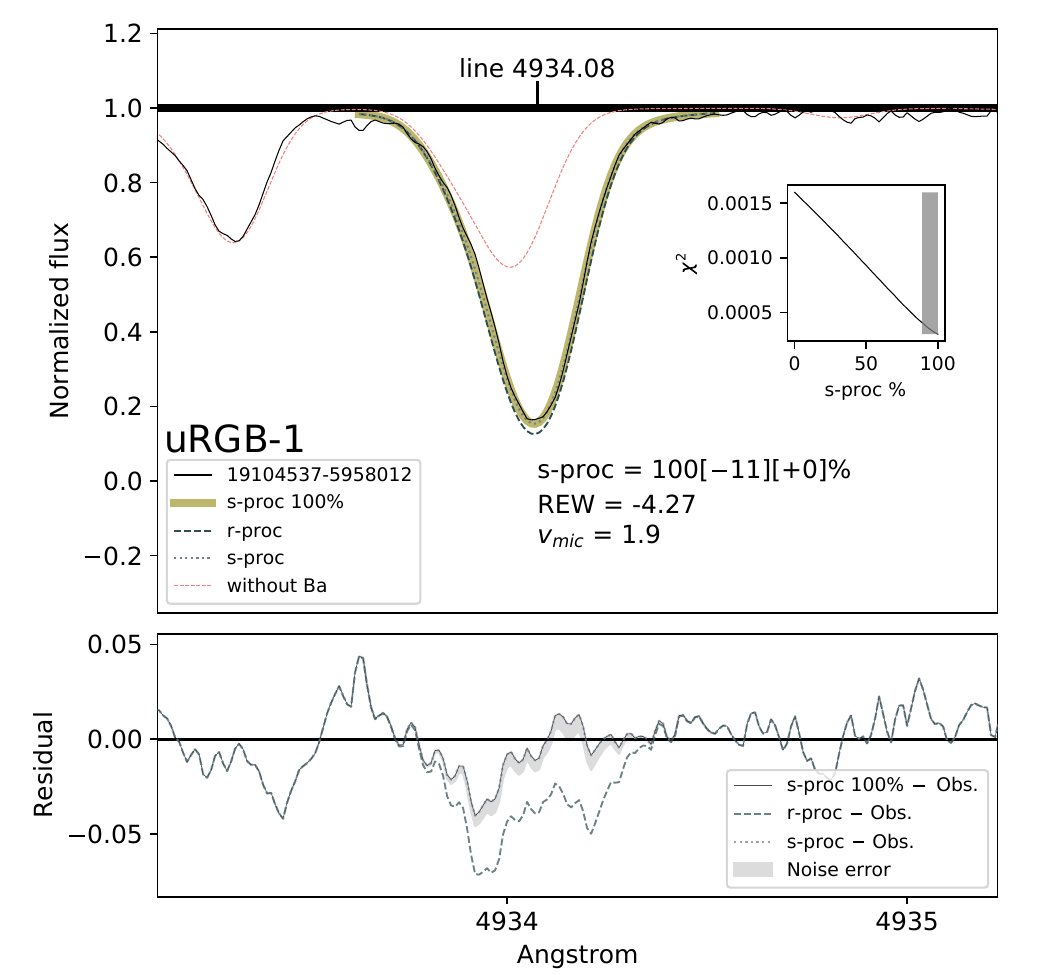}
    \includegraphics[width=0.32\linewidth]{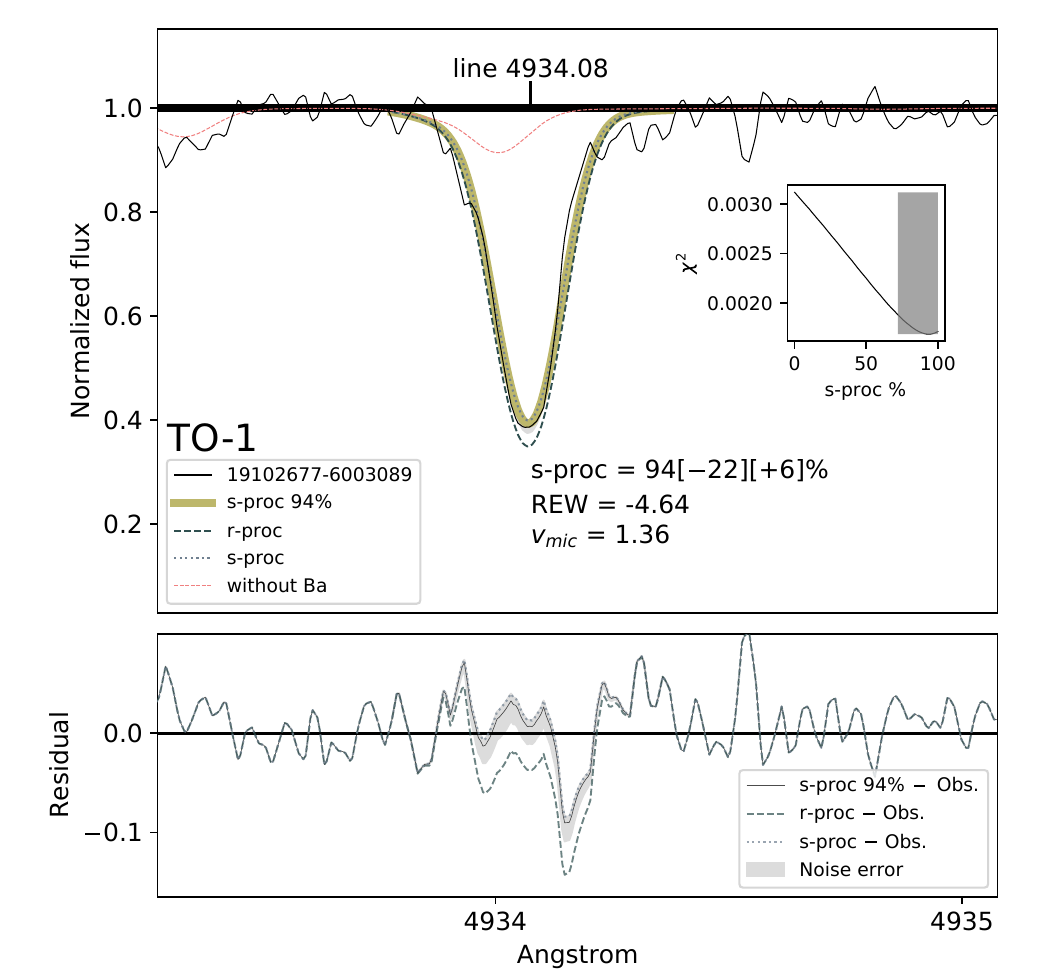}
    \includegraphics[width=0.32\linewidth]{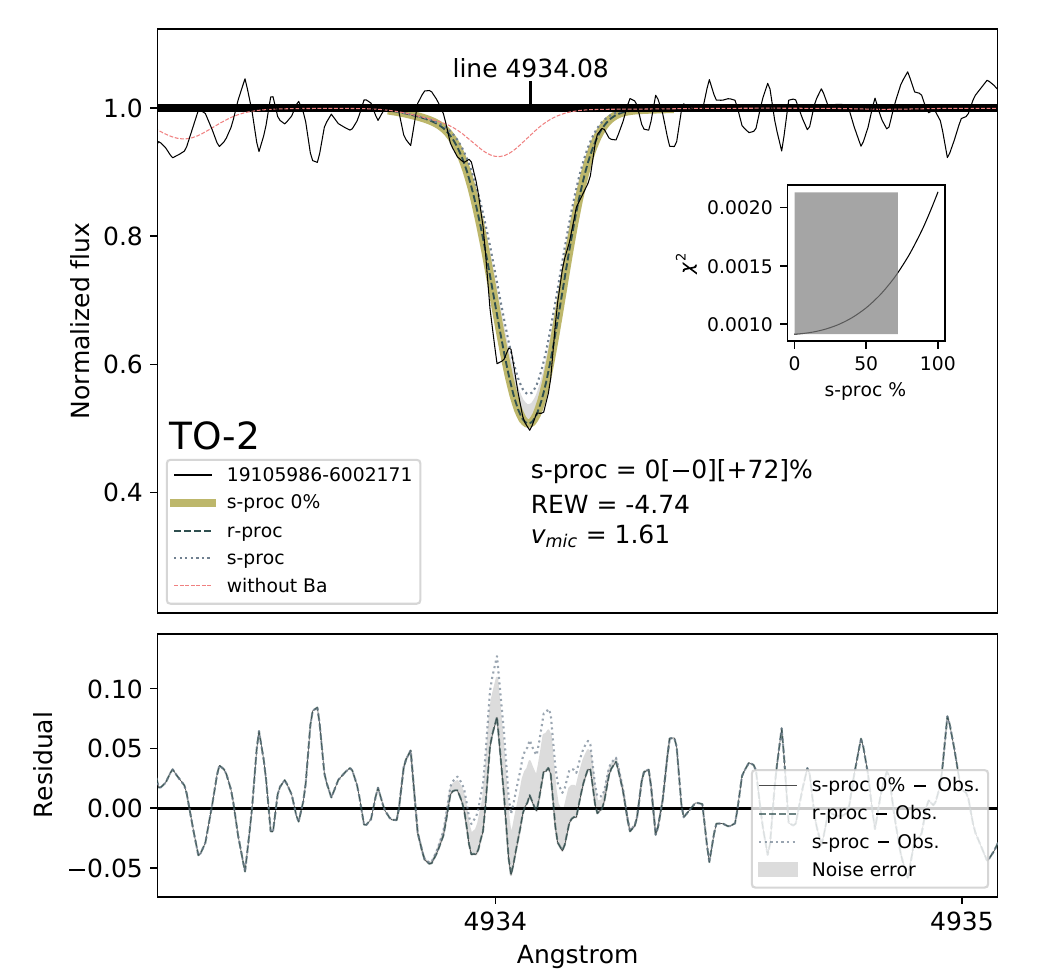}
    \caption{\tiny Fits of the 4934~\AA\ Ba resonance line.
    The stars to which each plot corresponds are indicated in the plots.
    The observational profiles are represented by the black lines. Synthetic profiles are represented by coloured lines.
    The synthetic profiles most compatible with the observational ones are represented by the thick olive lines; their s-process contributions are noted along with the errors related to the noise.
    Synthetic profiles related to s- and r-processes are represented by the dotted or dashed green lines according to the legends.
    A synthetic spectra without barium are represented by the dotted red lines.
    The probability for a given process to dominate the line profile shape is given by the $\chi^2$ in the inner plot.
    The shades in the inner plots cover errors of the s-process contribution.
    Residual plots are shown at the bottom panel, where the shades represent the areas covered by the fitting errors. 
    }
    \label{fig:TO1_spercent}
\end{figure*}

Figure~\ref{fig:TO1_spercent} shows the application of the method to the uRGB-1, 
TO-1, and TO-2 stars, as examples. 
The syntheses correspond to the A(Ba) listed in Table~\ref{tab:ba_abundances} with the s-process contribution percentage that fits better the observational profile. 
The plots also display line profiles corresponding to 0 (r-process) and 100\% s-process contributions. These are more easily  distinguished in the residual plots in the bottom panel.
The errors related to the flux variation due to spectral noise are noted within brackets. The inner plots display the $\chi^2$ minimisation used for fitting, where the shades cover the errors in brackets.
Only for the TO-2 star, the synthetic profile that best fits the observational line corresponds to the r-process.

The uncertainty on the isotopic ratios is primarily driven by the uncertainties in the stellar parameters that most significantly affect the line profiles, and it can be expressed as follows:

\begin{equation}
    \label{eq:error_iso}
    \Sigma_{iso} = \sqrt{ \sigma_{sn-iso}^2 + \sigma_{Ba-iso}^2 + \sigma_{T-iso}^2 + \sigma_{v-iso}^2 + \sigma_{[\mathrm{Fe/H}]}^{2} }
\end{equation}

\noindent
where $\sigma_{sn-iso}$ is the error related to noise, $\sigma_{Ba-iso}$ is the error induced by the A(Ba) errors, $\sigma_{T-iso}$ is the error induced by the \teff\ errors, $\sigma_{v-iso}$ is the error induced by the $v_{mic}$ errors, and $\sigma_{\mathrm{[Fe/H]}}$ is the error induced by [Fe/H] errors.
All these errors are individually listed in Table~\ref{tab:ba_abundances}. Details on how the various parameters influence the measurement of the isotopic ratio are provided in Sect.~\ref{sec:error:iso} in the Appendix.

\begin{table}
\caption{Errors of isotopic ratios.}
\label{tab:teff_errors}
\centering
\tiny 
\begin{threeparttable}
\begin{tabular}{l|ccccccccc}
\hline\hline
Parameter & 115 & 150 & 170 & 235 & 260\\
& [m\AA] & [m\AA] & [m\AA] & [m\AA] & [m\AA]  \\
\hline
A(Ba) & \multicolumn{5}{c}{ $\sigma$(A(Ba)) / 0.0035 }  \\
\teff & +12/$-16$ & +15/$-25$ & +16/$-27$ & +16/$-28$ & +16/$-29$\\
$v_{mic}$ & $\pm12$ & $\pm12$ & $\pm12$ & $\pm23$ & $\pm23$ \\
$\mathrm{[Fe/H]}$ & $\mp7$ & $\mp5$ & $\mp3$ & $\mp3$ & $\mp4$\\
\hline
\end{tabular}
\begin{tablenotes}
\item{} \textbf{Notes.} {The column headers express the EW of the resonance line at 4934~\AA. The unities of the quantities are expressed in terms of percentage of s-process contribution. 
Errors related to those of \teff\ are computed by deviating the true value by $\pm50$~K. 
Errors related to those of $v_{mic}$ correspond to variations of $\pm0.1$~km~s$^{-1}$. Errors related to those of [Fe/H] are computed by deviating the true value by $\pm0.1$~dex.
} 
\end{tablenotes}
\end{threeparttable}
\end{table}

\subsection{Hypothesis on the origin of the two TO stars}
\label{sec:MSTO}
Although the primary aim of this paper is not to analyse the barium abundance and isotopic composition of NGC~6752, but rather to provide calibration relations for microturbulence, we cannot avoid commenting on the TO stars.
One of the most intriguing hypotheses is that the lower barium abundance, combined with an isotopic ratio fully dominated by the r-process, could point to a first-generation star within the framework of multiple stellar populations in globular clusters.  
In fact, one of the most distinctive characteristics of GCs is the presence of multiple stellar populations—a phenomenon observed in nearly all Galactic GCs. The prevailing theory suggests that GCs initially formed a first generation (FG) of stars from pristine gas, meaning gas with the original, unprocessed chemical composition. A subset of these stars—commonly referred to as polluters—then enriched the intra-cluster medium with elements synthesised during their lifetimes. This enriched material was subsequently mixed with the remaining pristine gas, leading to the formation of a second generation (SeG) of stars. These different stellar populations can be identified through their distinct chemical signatures, particularly in the abundances of hot H-burning elements such as oxygen, sodium, magnesium, and aluminium. \citet{Schiappacasse-Ulloa2022} presented evidence that intermediate-mass asymptotic giant branch (AGB) stars (4–8 M$_\odot$) contributed to this enrichment in NGC~6752, which may contribute with some degree of s-process elements, varying their yields according to the mass and metallicity of the star. However, \citet{Schiappacasse-Ulloa2023} found no significant differences in the abundances of Ba or other s-process elements between the stellar populations, despite observing a substantial spread in Ba. 
Table~\ref{tab:BasicInfo} lists in the last column the stellar population membership—first generation (FG) or second generation (SeG)—as determined by Na abundances from \citet{Schiappacasse-Ulloa2025}. Except for TO-2, none of the stars exhibit statistically significant differences in either A(Ba) or isotopic composition, regardless of their population. This weakens the hypothesis of a systematic difference in isotopic ratios between the two generations.
The similar Ba abundances observed in both FG and SeG stars support the findings of \citet{Schiappacasse-Ulloa2023}, and also suggest that the anomalous Ba abundance in TO-2 is unlikely to be associated with the multiple population phenomenon. 

Another possible hypothesis is that TO-2 is not a cluster member and was accreted at a later stage. 
However, its metallicity is consistent with that of the other TO stars, and its astrometric and kinematic properties strongly support its membership.

This star will be discussed in greater detail in Paper~III, along with the other globular clusters observed as part of the {\it Gaia}-ESO Survey and already presented in \citet{Schiappacasse-Ulloa2025}.

\section{Discussion and Conclusions}
\label{sec:conclusions}

To the present date, many CEMP stars are found to be enhanced in both r- and s-processes (usually labelled as CEMP-rs).
Hypothetical scenarios explaining how such stars ended up with that chemical composition pattern consist of either pollution \citep[of the primordial cloud or the already formed star system, e.g. ][]{beers2005ARA&A..43..531B,jonsell2006A&A...451..651J,hansen2016A&A...588A...3H,gull2018ApJ...862..174G}
or an intermediate (i-) speed nucleosynthesis process \citep{cowan1977ApJ...212..149C}. The latter has been theoretically investigated \citep[e.g.][]{choplin2021A&A...648A.119C, goriely2021A&A...654A.129G, choplin2022A&A...667A.155C, martinet2024A&A...684A...8M, choplin2024A&A...684A.206C, Denissenkov2021MNRAS.503.3913D}. However, no a star with a chemical pattern consistent with the predictions of the i-process was identified yet.

The most rigorous method used to classify CEMP stars into r-, s-, or rs-process subgroups consists of comparing their heavy element abundances with theoretical nucleosynthesis predictions \citep[e.g.][]{gull2018ApJ...862..174G, sbordone2020A&A...641A.135S, da_silva2025A&A...696A.122D}.  
However, the method is prone to inaccurate results from line modelling using classical 1D model atmospheres (under LTE or NLTE) because in CEMP stars (which are mostly RGB) spectral lines are often very intense and also may be severely blended with molecular features.
In this context, a competitive alternative method is the determination of isotopic ratios of heavy elements. 
The barium resonance lines may serve as an effective indicator, as these are shaped according to the dominating nucleosynthesis process.
The s-process produces even isotopes (134, 136, and 138) in greater quantities than the r-process, which mostly produces odd isotopes (135 and 137); see Table~\ref{tab:ratios}.
Since the latter mostly influence the line profile wings, whereas the former mostly influence the line core, their line profiles are certainly distinguishable (see Fig.~\ref{fig:comparison_profiles}) in spectra of sufficient quality \citep[e.g.][]{mashonkina2006A&A...456..313M,Gallagher2020A&A...634A..55G,cescutti2021A&A...654A.164C}.
Additionally, the i-process shapes a line that is almost identical to that of the r-process (blue shade and solid line in Fig.~\ref{fig:comparison_profiles}, respectively). 
Therefore, in this spectral feature, a star where both s- and r-processes dominated should be distinguishable from one where the i-process was dominant.

In principle, it should be somewhat straightforward to model Ba resonance lines via spectral synthesis.
However, these are strong in CEMP stars; therefore, their isotopic diagnoses made without considering ad hoc parameter calibrations are unreliable, given that 1D model atmospheres are unsuitable to lines of such strength.
In order to take advantage of this tool of great potential, and to promote its use, we analysed the behaviour of barium lines in metal-poor stars along several evolutionary stages spanning from the turn-off to the upper RGB.
For this, we use stars in the cluster NGC~6752 (Fig. \ref{fig:kiel}), whose A(Ba) abundances are reasonably homogenous \citep[$\pm0.1$~dex, e.g.][]{Schiappacasse-Ulloa2023}.
This characteristic allows us to fix the abundance in the spectral synthesis to calibrate microturbulence quantities according to the line strengths.
Multiple stars were included at each evolutionary stage (Figs.~\ref{fig:cmd} and \ref{fig:kiel}) to ensure that our conclusions are not based on atypical cases.

Using turn-off stars as A(Ba) reference --we fixed the same abundance for all cluster stars, 
we first calibrated the microturbulence (\vcalibsub) to several subordinate lines $\lambda$5853, $\lambda$6141, $\lambda$6496~\AA.
We did not find substantial difference with $v_{mic}$ computed from Fe lines under 1D LTE, except for SG stars, where \vcalibsub\ is lower by $\sim$0.40~km~s$^{-1}$ (see Fig~\ref{fig:vmic_teff}).
On the other hand we find that \vcalibsub\ is compatible with the 3D model-based microturbulence provided by the \teff-\logg\ dependent relation of \cite{dutra-ferreira2016A&A...585A..75D}, here reproduced in Eq~\ref{eq:vmic3D}.
We recommend the use of that relation for the determination of the barium abundance using any of the subordinate lines. This may facilitate the determination of atmospheric parameters in CEMP stars, as in kind of stars determining $v_{mic}$ becomes challenging due to the scarcity of weak and saturated Fe lines.

We assumed that the dominant nucleosynthesis process in the TO-1 star (s-process, see Fig.~\ref{fig:TO1_spercent}) is also dominant in the other cluster stars.
This way, fixing the isotopic ratios (and also fixing A(Ba) to that of the TO-1 star), we calibrated the microturbulence for the resonance line at 4934~\AA. 
Related tests are shown in Fig.~\ref{fig:vmic_teff_4934}, where results of the correct hypothesis (A(Ba) = 1.15~dex and 100\% s-process contribution, left panel) and incorrect hypotheses (100\% r-process contribution and A(Ba) = 1.19 and 0.39~dex, centre and right panels, respectively) are shown.
We find that the microturbulence adapted to the resonance line at $\lambda$4924~\AA\ (\vcalibres) is substantially higher (by $\sim$0.4~km~s$^{-1}$) than that from Fe lines for stars in the upper part of the RGB. 
When this difference is not considered, a correct diagnosis of 100\% s-process may be switched to an incorrect one of $\sim$100\% r-process, for example; see quantities related to $v_{mic} \pm$0.1~km~s$^{-1}$ in Table~\ref{tab:teff_errors}.
Figure~\ref{fig:vcalib_REW} shows our \vcalibres\ values as function of the EW and REW compared with $v_{mic}$ from Fe lines and from the 3D model-based relation.
We provide a polynomial fit in Eq.~\ref{eq:pol}, and, alternatively the quantities accompanying the dashed line in the plot, for a practical use of these adapted microturbulence.
Regarding our 1D LTE barium abundance scale, it is higher than 1D NLTE by $\sim$0.2~dex, as shown in Fig.~\ref{fig:vmic_var_NLTE}. However, we note that according to 3D NLTE prototype models\footnote{\url{https://www.chetec-infra.eu/3dnlte/abundance-corrections/barium/}},
1D LTE barium abundances of typical TO and RGB stars are very close 3D NLTE, see example in Fig.~\ref{fig:3Dcors}.
Our stars are about 12~Gyr old and none of them has entered in the Asymptotic Giant Branch phase, therefore our microturbulence calibrations are free of biases from chromospheric activity effects that may underlay the \textit{Barium puzzle} (see Sect.~\ref{sec:activity}).

A by-product of our calibration work is the determination of barium abundances and isotopic ratios of the cluster stars; these are listed in Table~\ref{tab:ba_abundances} along with the errors induced by the errors of the atmospheric parameters, individually.
Table~\ref{tab:teff_errors} provides estimates of the errors for their practical use. 
Our isotopic ratio determinations are given in terms of the s-process contribution from 0 to 100\% assuming that the total barium abundance is composed solely by r- and s-process products.
For spectra of S/N $\sim 50$, such as those of our TO stars, the High-Resolution Multi-Object Spectrograph (HRMOS) instrument with resolution $R \approx 80\,000$ will nearly double the precision of the isotopic ratio determination \citep[][Fig~32]{magrini2023arXiv231208270M};
our $v_{mic}$ calibrations are precisely made to take full advantage of the instrument products.

\begin{acknowledgements}
R.E.G., L.M., J.S.U. and S.R. thank INAF for the support (Large Grants EPOCH and WST), the Mini-Grants Checs (1.05.23.04.02), and the financial support under the National Recovery and Resilience Plan (NRRP), Mission 4, Component 2, Investment 1.1, Call for tender No. 104 published on 2.2.2022 by the Italian Ministry of University and Research (MUR), funded by the European Union – NextGenerationEU – Project ‘Cosmic POT’ Grant Assignment Decree No. 2022X4TM3H by the Italian Ministry of the University and Research (MUR).
    J.S.U. thanks INAF for its support through the Mini-Grant (1.05.24.07.02). The 3D NLTE corrections used in this work were provided by the ChETEC-INFRA project (EU project no. 101008324), task 5.1.
    Use was made of the Simbad database, operated at the CDS, Strasbourg, France, and of NASA’s Astrophysics Data System Bibliographic Services. 
    This publication makes use of data products from the Two Micron
    All Sky Survey, which is a joint project of the University of
    Massachusetts and the Infrared Processing and Analysis
    Center/California Institute of Technology, funded by the National Aeronautics and Space Administration and the National Science Foundation.
    This research used Astropy (\url{http://www.astropy.org}) a community-developed core Python package for Astronomy \citep{astropy:2018}.
    This work presents results from the European Space Agency (ESA)
    space mission Gaia. Gaia data are processed by the Gaia Data Processing and Analysis Consortium (DPAC). Funding for the DPAC is provided by national institutions, in particular the institutions participating in the Gaia MultiLateral Agreement (MLA). The Gaia mission website is \url{https://www.cosmos.esa.int/gaia}. The Gaia archive website is \url{https://archives.esac.esa.int/gaia}. Based on spectroscopic data obtained with ESO
Telescopes at the La Silla Paranal Observatory under programmes 188.B-3002,
193.B-0936, and 197.B-1074, available at \url{https://archive.eso.org/scienceportal/home?data_collection=GAIAESO%5C&publ_date=2020-12-09}.
\end{acknowledgements}

\bibliographystyle{aa.bst}

\bibliography{Faint2}

\newpage
\begin{appendix} 

\section{Effective temperature determinations}

In this section, we report the effective temperature determinations of our star sample using the methods described in Sec.~\ref{sec:parameters}.
\begin{table}[!h]
\caption{Effective temperature determinations}
\label{tab:teffs}
\centering
\tiny 
\begin{threeparttable}
\begin{tabular}{lccccccccc}
\hline\hline
Star & \teffa & \teff($B_p-R_p$) & \teff($G-B_p$)  \\
\hline
Pavo uRGB-1 & $4520 \pm 45$ & $4457 \pm 74$ & $4502 \pm 75$ \\
Pavo uRGB-2 & $4483 \pm 21$ & $4418 \pm 74$ & $4498 \pm 76$ \\
Pavo ocRGB-1 & $4654 \pm 50$ & $4510 \pm 74$ & $4502 \pm 76$\\
Pavo ocRGB-2 & $4724 \pm 68$ & $4649 \pm 73$ & $4659 \pm 76$\\
Pavo bRGB-1 & $5238 \pm 109$ & $5116 \pm 74$ & $5126 \pm 80$\\
Pavo bRGB-2 & $5054 \pm 148$ & $5107 \pm 74$ & $5286 \pm 82^\dagger$\\
Pavo SG-1 & $5447 \pm 39$ & $5426 \pm 73$ & $6443 \pm 94^\dagger$ \\
Pavo SG-2 & $5430 \pm 118$ & $5322 \pm 73$ & $5312 \pm 84$\\
Pavo SG-3 & $5381 \pm 105$ & $5282 \pm 73$ & $5775 \pm 87^\dagger$\\
Pavo SG-4 & $5326 \pm 80$ & $5297 \pm 73$ & $5531 \pm 84$\\
Pavo TO-1 & $6189 \pm 197$ & $6038 \pm 76$ & $6068 \pm 92$\\
Pavo TO-2 & $6274 \pm151$ & $6176 \pm 76$ & $6427 \pm 98^\dagger$\\
Pavo TO-3 & $6145 \pm 359$ & $5957 \pm 100^\dagger$ & --- \\
Pavo TO-4 & $6285 \pm 202$ & $6043 \pm 65$ & $6070 \pm 92$\\
\hline
\hline
\end{tabular}
\begin{tablenotes}
\item{} \textbf{Notes.} {The symbol ($^\dagger$) indicates a quantity not considered for determining the average \teff\ in Table~\ref{tab:atmo_param}. 
} 
\end{tablenotes}
\end{threeparttable}
\end{table}

\section{Barium abundances and microturbulence of resonance lines in different evolutionary stages}

\begin{figure}
    \centering
    \includegraphics[width=0.9\linewidth]{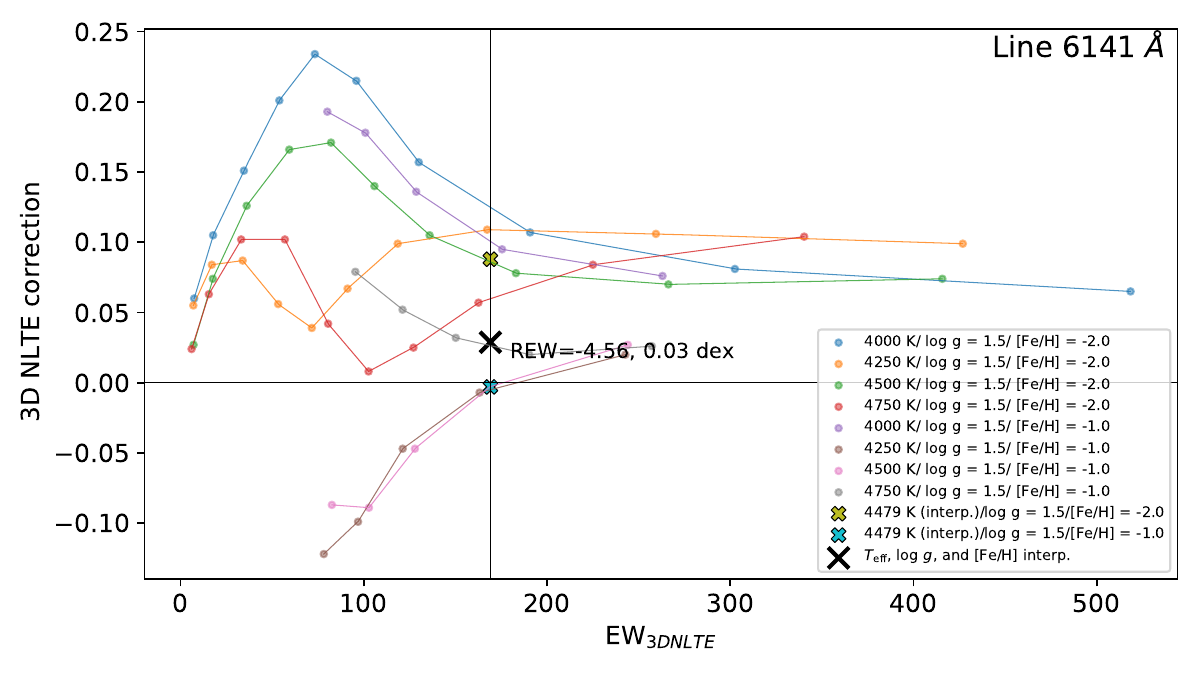}
    \caption{\tiny Interpolation of 3D NLTE corrections of the line at $\lambda$6141~\AA\ for the star uRGB-2. Circles represent values listed in the original tables, the parameters of which are given in the legends. 
    Lines, coloured the same as the circles, are linear interpolations.
    The crosses represent values interpolated in \teff\ and \logg. The black cross displays the interpolated value in \teff, \logg, and [Fe/H].
    The REW of the line and the 3D NLTE corrections are noted in the plot.
    }
    \label{fig:3Dcors}
\end{figure}

It is clear that for the uRGB and ocRGB stages, \vcalibres\ is significantly higher than \vlte, \vnlte, and \vfit\ for the 100\% s-process contribution (left panel in Fig.~\ref{fig:vmic_teff_4934}). On the other hand, \vcalibres\ is compatible with \vlte, \vnlte, and \vfit\ for full r-process profiles (right panel in Fig.~\ref{fig:vmic_teff_4934}).
This is a clear evidence that using microturbulence from Fe lines (either \vlte, \vnlte, or \vfit) to derive isotopic ratios from modelling Ba resonance line profiles, may likely provide misleading diagnoses. 
The barium in these stars has been determined to be produced mostly by the s-process according to the analysis in Sect.~\ref{sec:calib_subordinate}.
Therefore, the offsets on the left panel of Fig.~\ref{fig:vmic_teff_4934} quantify the corrections that $v_{mic}$ of Fe lines require to be used for the Ba resonance line at 4934~\AA, i.e. +0.4~km~s$^{-1}$. 
One star (ocRBG-1) shows a slightly lower difference. Possibly, this is a small bias related to the method used to determine \vlte, as we obtain for this star a \vfit\ compatible with \vlte\ of the other three stars.  
The stars uRGB-2 and ocURGB-1 have $\sigma_{sn-iso} = 0$ in Table~\ref{tab:ba_abundances}. This is because no isotopic ratio combination in between the s- and the r-process are able to fit the observational profile better than that of the s-process. Only decreasing the abundance of these stars, (by increasing their \vcalibsub) would be compensated by a decrease of the s-process contribution percentage.
This would require \vcalibsub\ higher than those in Table~\ref{tab:ba_abundances}, thus higher than \vlte\ as well.
As we explain in Sect.~\ref{sec:calib_subordinate}, it does not seem plausible that our programme stars (except TO-2) have Ba from the r-process because their \vcalibsub\ would need to be as high as 2-3~km~s$^{-1}$ (see Fig.~\ref{fig:vmic_teff}).
However, it could still be argued that contributions of about 70-90\% s-process are possible, as other stars in our sample present. 
To explore this possibility, we interpolated 3D NLTE corrections\footnote{\url{https://www.chetec-infra.eu/3dnlte/abundance-corrections/barium/}} computed by the code Linfor3D \citep{Linfor3D}.
We obtain a minor value of +0.03~dex from the lines 6141 and 6496~\AA, which support our  A(Ba) determinations under 1D LTE and their corresponding \vcalibsub.
Corrections towards lower A(Ba) are suitable for stars with both [Fe/H] $= -1$~dex and weaker Ba lines (e.g. EW $\sim 100$ or REW $\sim -4.9$); see Fig.~\ref{fig:3Dcors}.

\begin{table}
\centering
\tiny
\caption{Ba resonance line at 4934~\AA\ with isotopic and hyperfine structure splitting}
\begin{tabular}{cccccc}
\hline\hline
Isotope & Wavelength [\AA] & $F_\mathrm{low}$ & $F_\mathrm{up}$ & strength & log $gf$\\
\hline
$^{134}$Ba & 4934.075 &   &   & 1   & $-0.172$   \\ 
\\
\multirow{4}{*}{$^{135}$Ba} & 4934.047 & 1 & 1 & 0.0625 & $-1.376$\\ 
 & 4934.036 & 1 & 2 & 0.3125 & $-0.677$ \\ 
 & 4934.105 & 2 & 1 & 0.3125 & $-0.677$\\ 
 & 4934.094 & 2 & 2 & 0.3125 & $-0.677$\\ 
\\
$^{136}$Ba & 4934.076 &   &   & 1  &$-0.172$    \\ 
\\
\multirow{4}{*}{$^{137}$Ba} & 4934.043 & 1 & 1 & 0.0625 &  $-1.376$\\ 
& 4934.031 & 1 & 2 & 0.3125 & $-0.677$ \\ 
& 4934.109 & 2 & 1 & 0.3125 & $-0.677$\\ 
 & 4934.097 & 2 & 2 & 0.3125 & $-0.677$\\ 
\\
$^{138}$Ba & 4934.077 &   &   & 1   & $-0.172$   \\ 
\hline
\end{tabular}
\tablefoot{$F$ is the hyperfine interaction quantum number resulting from the coupling between the nucleus spin and the orbital angular momentum.  The strength is relative to the isotope considered.}
\label{tab:tab1}
\end{table}

\section{The differences between TO-1 and TO-2}
\begin{figure*}
    \centering
    \includegraphics[width=0.20\linewidth]{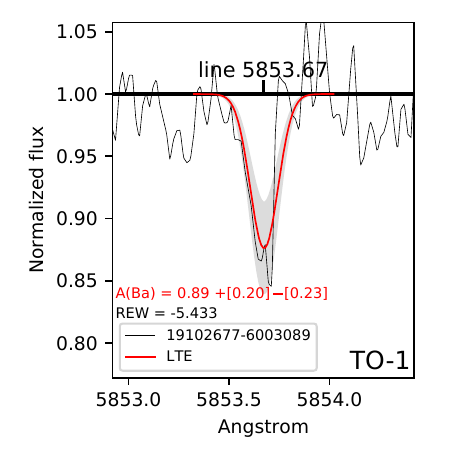}
    \includegraphics[width=0.20\linewidth]{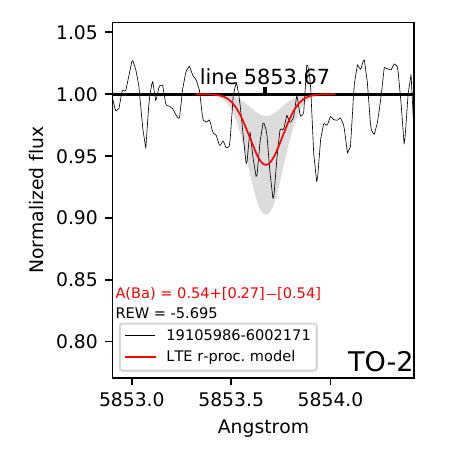}
    \includegraphics[width=0.20\linewidth]{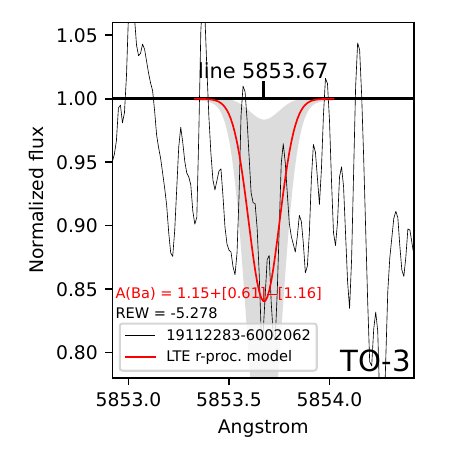}
    \includegraphics[width=0.20\linewidth]{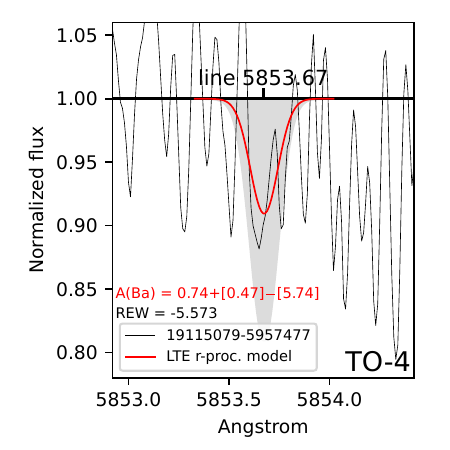}
    \\
    \includegraphics[width=0.20\linewidth]{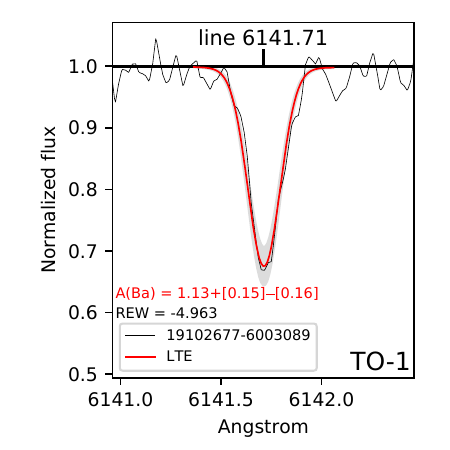}
    \includegraphics[width=0.20\linewidth]{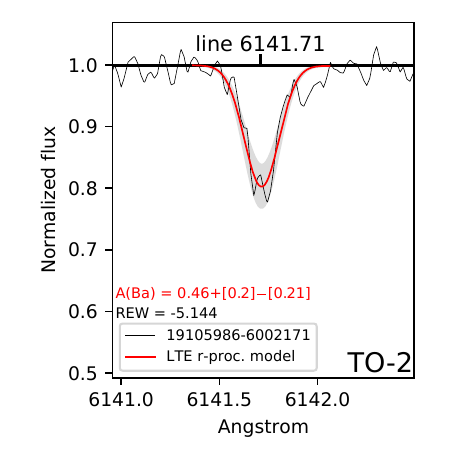}
    \includegraphics[width=0.20\linewidth]{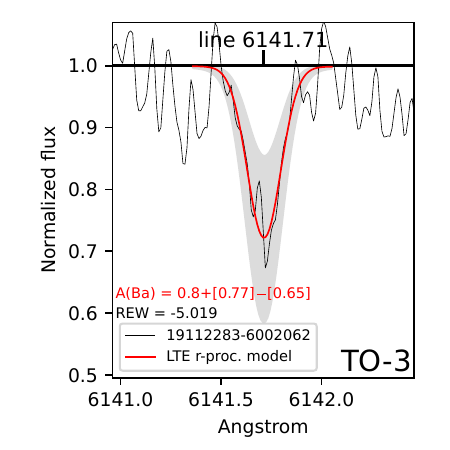}
    \includegraphics[width=0.20\linewidth]{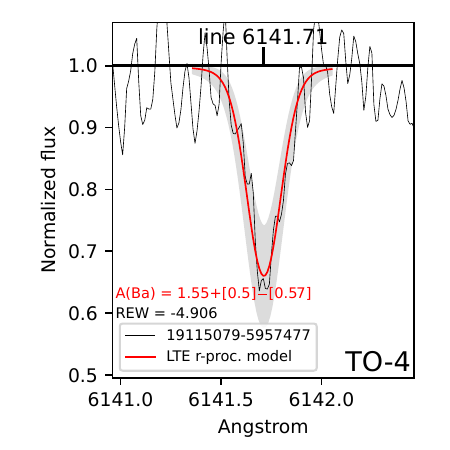}
    \\
    \includegraphics[width=0.20\linewidth]{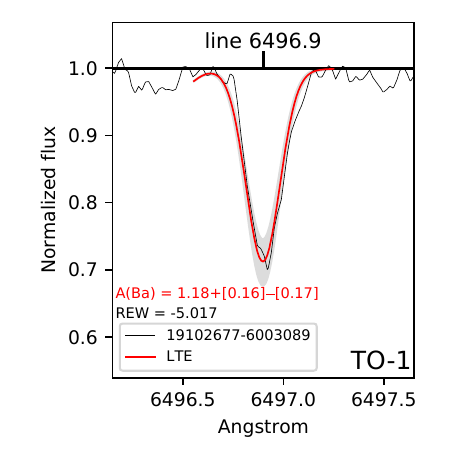}
    \includegraphics[width=0.20\linewidth]{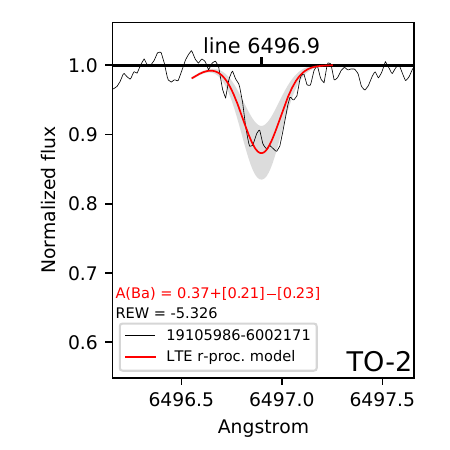}
    \includegraphics[width=0.20\linewidth]{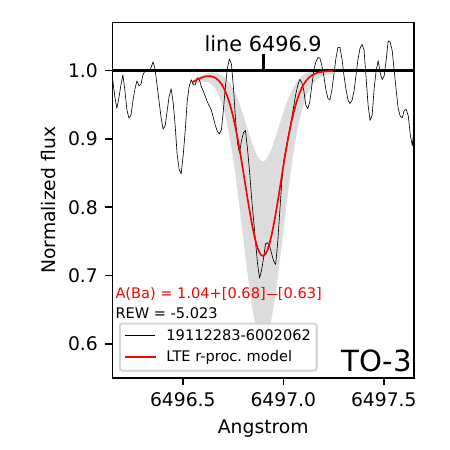}
    \includegraphics[width=0.20\linewidth]{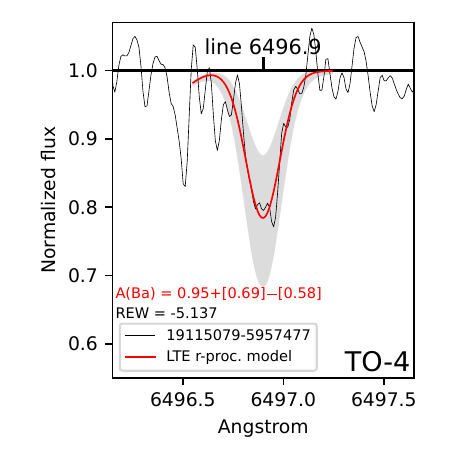}
    \caption{\tiny Fits of the subordinate lines of the TO  stars. Left and right panels show spectra of the TO-1, TO-2, TO-3, and TO-4 stars, respectively; these share the same scales. 
    Red lines represent synthetic line profiles under LTE, its associated abundance are noted in red with errors between brackets. 
    Shade areas represent fitting errors related to the noise, which are noted within brackets.
    }
    \label{fig:TO2_subordinate}
\end{figure*}

Figure~\ref{fig:TO2_subordinate} shows line fits of all TO stars. 
From top to bottom, each row shows the lines 5853, 6141, and 6496~\AA, respectively. 
We note that the lines of the TO-2 star are visibly smaller than those of the other stars; REW are annotated in the plots. 
Since the atmospheric parameters of these stars are very similar, the  line strengths of TO-2 support its relative lower abundance.
The low Ba abundance of the TO-2 star is also noted by \cite{Schiappacasse-Ulloa2023}.

\section{Detailed estimate on the Ba isotopic ratio uncertainties }
\label{sec:error:iso}

\begin{figure*}
    \centering
    \includegraphics[width=0.32\linewidth]{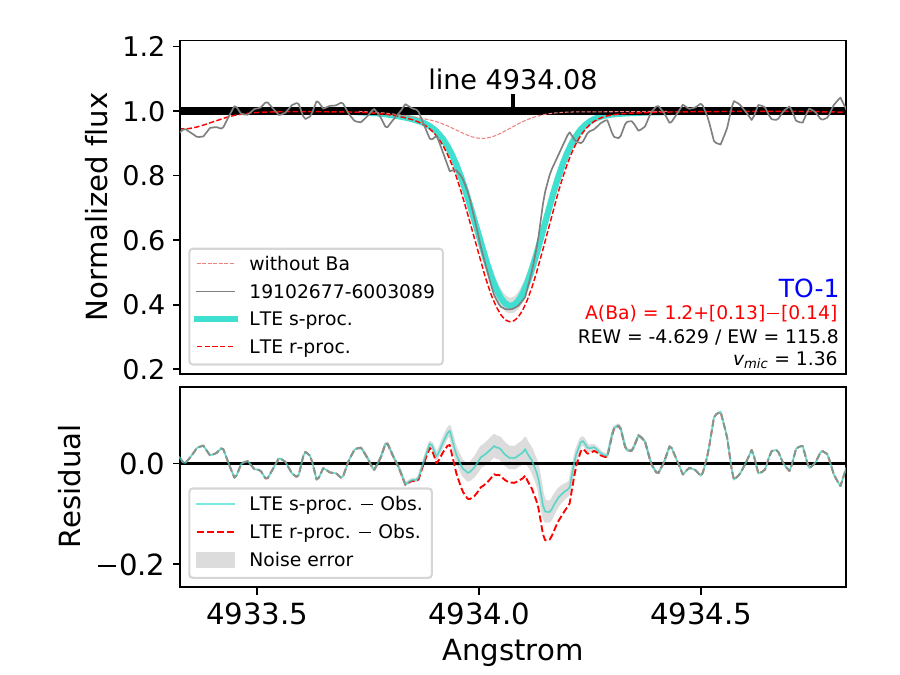}
    \includegraphics[width=0.32\linewidth]{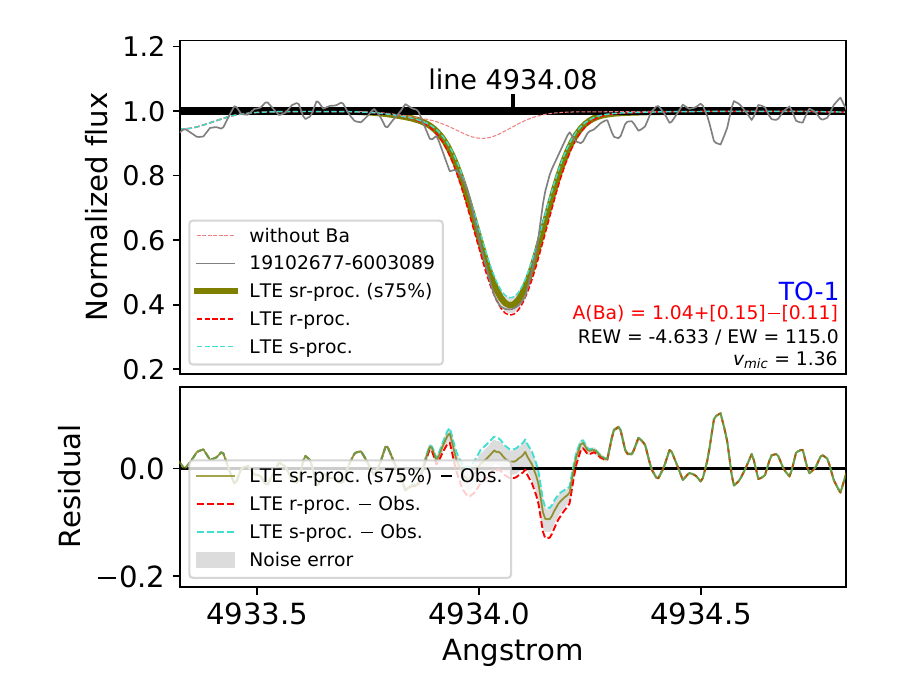}
    \includegraphics[width=0.32\linewidth]{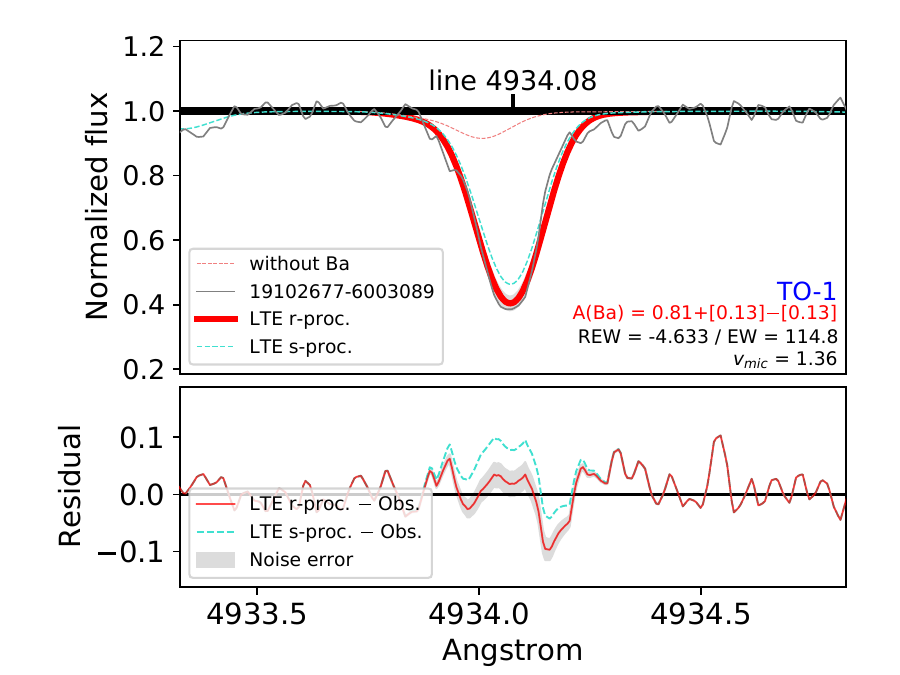} 
    \includegraphics[width=0.32\linewidth]{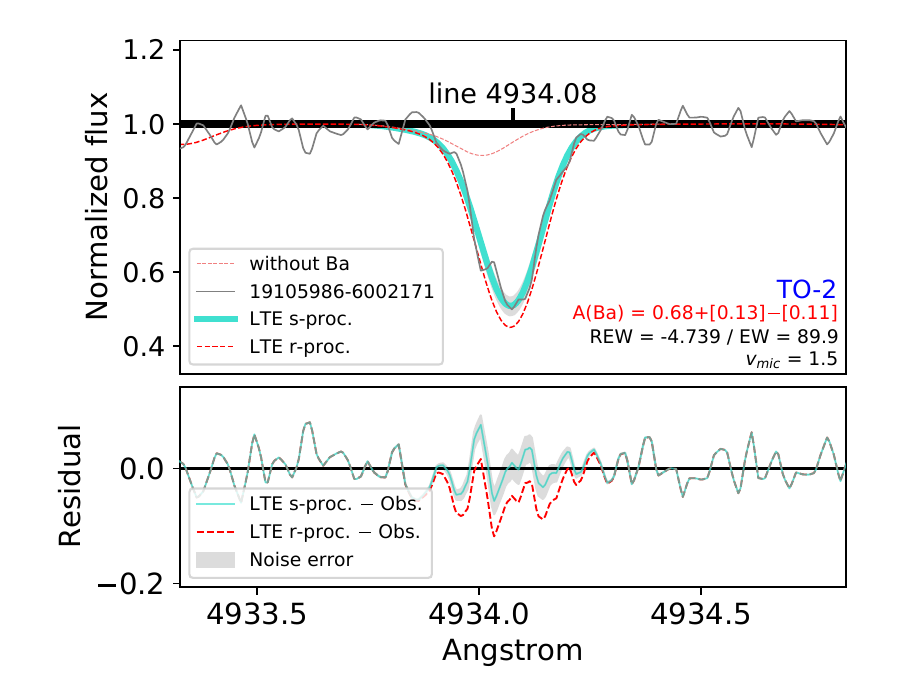} 
    \includegraphics[width=0.32\linewidth]{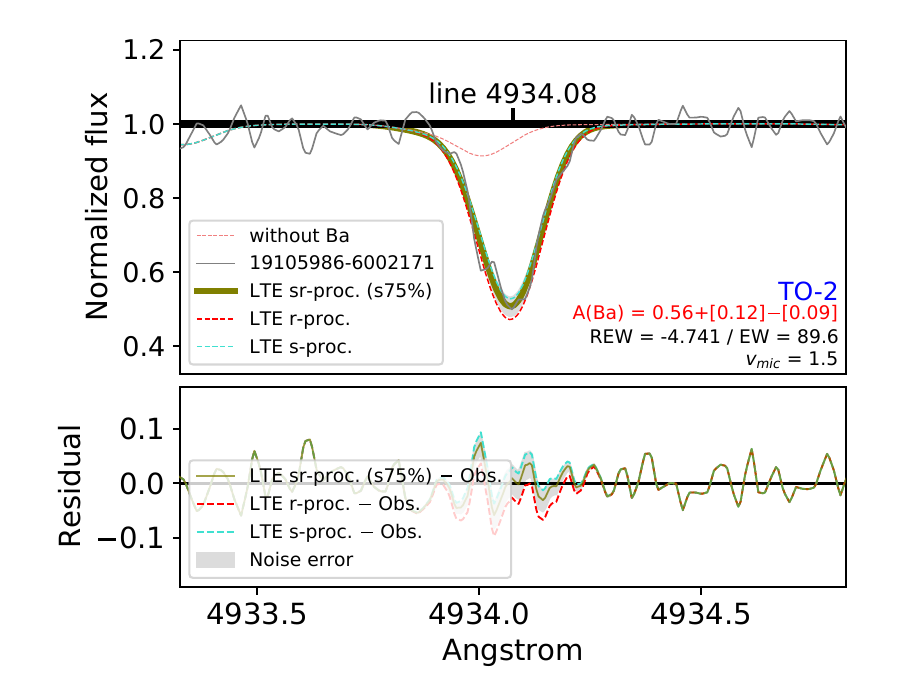} 
    \includegraphics[width=0.32\linewidth]{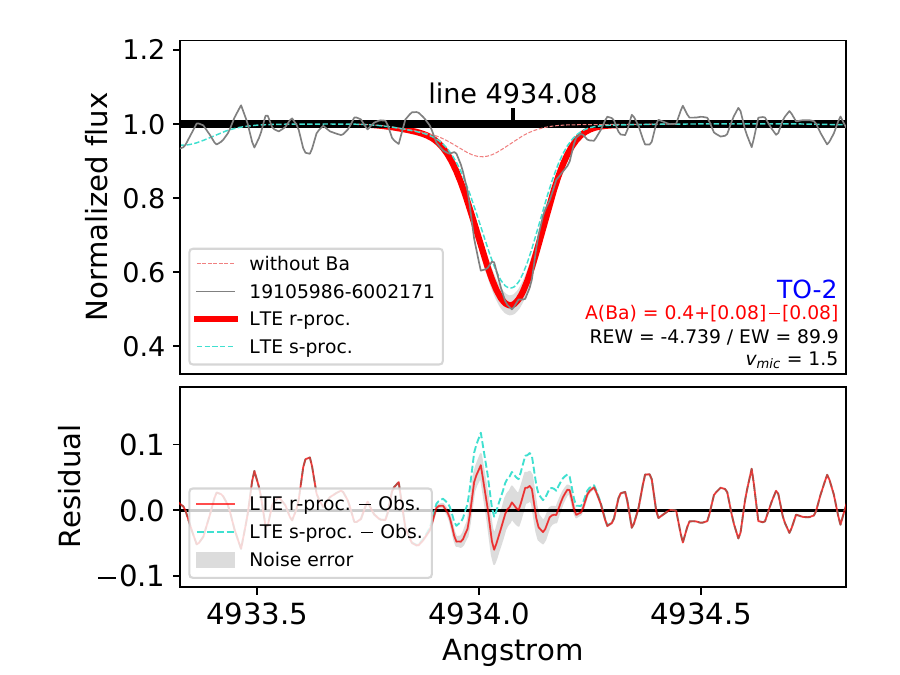} 
    \caption{\tiny Fits of the resonance line 4934~\AA\ of the TO stars. Top and bottom panels correspond to the TO-1 and TO-2 star, respectively. 
    Left, middle, and right panels show fits of 100, 75, and 0\% s-process profiles, respectively. For comparison, every panel shows line profiles of s- (cyan dashed line) and r-process (red dashed line) with an abundance equal to that determined by the fits (characters in red). 
    Residuals of the fits are shown below the main plots. Fitting errors related to the noise are covered by shades.
    }
    \label{fig:TO2_resonance}
\end{figure*}

First we examine $\sigma_{Ba-iso}$.
In Fig.~\ref{fig:TO2_resonance} we provide examples of the degeneracy of barium abundance with isotopic ratios for the resolution of our spectra.
We test 100, 75, and 0\% s-process contributions letting A(Ba) vary freely.
For TO-1, we obtain A(Ba) values compatible with that determined from subordinate lines ($1.15 \pm 0.03$~dex) for 100 and 75\% s-process profiles (top left and middle panels); whereas the abundance from the r-process (top right panel) is certainly too low.
For TO-2, only the A(Ba)  from the 0\% s-process is compatible with the values determined from the subordinate lines ($0.39 \pm 0.02$~dex).
r-process profiles are wider than s-process ones due to the prominence of odd isotopes, which are split towards the line wings; see Fig.~\ref{fig:comparison_profiles}.
For spectra of $R \sim 45\,000$, r-process profiles appear deeper than s-process ones (compare the cyan and red profiles in Fig.~\ref{fig:TO2_resonance}, for example), thus these yield lower abundances when fitted to observational lines.
These results indicate that an abundance variation of about +0.35~dex\footnote{Similar values are found in both dwarfs and giants.} is able to change the diagnosis of isotopic ratio from r- to s-process (i.e. from 0 to 100\% s-process). 
Assuming a linear relation, a variation of $\pm1$\% of s-process contribution is associated to $\pm0.0035$~dex; therefore $\sigma_{Ba-iso}$ (in percentage) is given by $\sigma_{sn}$ between 0.0035.
Since our estimate of $\Sigma_{iso}$ include the effects of \teff\ and $v_{mic}$ errors separately, we determine $\sigma_{Ba-iso}$ only using the dispersion of the abundances of the subordinate lines (i.e. $\sigma_{sn}$).

We estimate $\sigma_{T-iso}$ using a synthetic grid of the line 4934~\AA\ with parameters representing the defined evolutionary stages (TO, SG, bRGB, ocRGB, and uRGB) and assuming s-process contribution of 50\%. We varied \teff\ of the spectra of simulated stars by $\pm50$~K with respect to the values of the grid. 
We obtain the percentage variations in Table~\ref{tab:teff_errors}.
$\sigma_{T-iso}$ affect more seriously the isotope ratios diagnosis than any other source. As the quantities in the table show, a \teff\ change as low as $50$~K may change the contribution of the s-process by up to $30$\%.
Considering that typical \teff\ uncertainties of metal-poor stars in spectroscopic surveys are of $\sim$100~K, and that custom spectroscopic methods are prone to bias \teff\ by more than 100~K in metal-poor stars \citep[e.g. Figs.~2, 11, and A.5 in][ respectively]{giribaldi2023A&A...673A..18G, giribaldi2023A&A...679A.110G, giribaldi20252025A&A...698A..11G},
the accurate determination of \teff\ becomes paramount in the diagnosis of isotopic ratios.

\begin{figure*}
    \centering
    \includegraphics[width=0.33\linewidth]{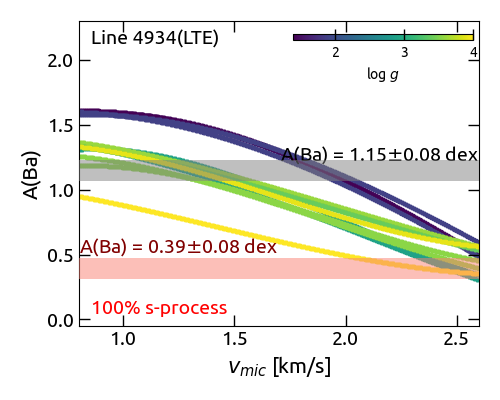}
    \includegraphics[width=0.33\linewidth]{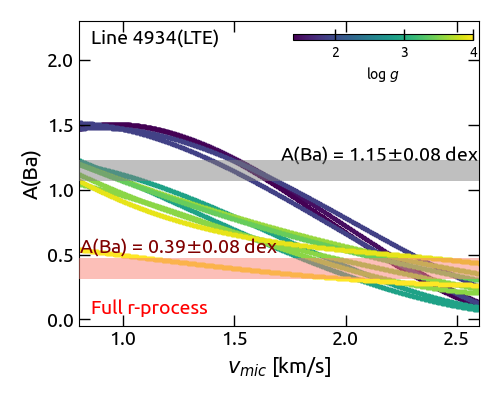}
    \caption{\tiny Similar to Fig.~\ref{fig:vmic_var} for the line at 4934~\AA. From left to right panels, the isotopic ratios assumed for the synthesis correspond to 100 and 0\% s-process contributions.
    Gray and pink shades cover A(Ba) of the TO-1 and TO-2 stars, respectively, including dispersions. The dispersions of $\pm0.08$ ~dex are $\sigma_{sn}$ and $\sigma_T$ added in quadrature (see Table~\ref{tab:ba_abundances}).}
    \label{fig:vmic_var_4934}
\end{figure*}

Regarding $\sigma_{v-iso}$,
an estimate of its impact in the s-process contribution percentage is obtained via the impact of A(Ba).
In Fig.~\ref{fig:vmic_var_4934} we show the change of A(Ba) with $v_{mic}$ in the line 4934~\AA\ for all the evolutionary stages, where the abundance of the cluster is set to be A(Ba) = $1.15 \pm 0.08$~dex (gray shade).
Using the fraction defined above for $\sigma_{sn}$, we can make the division of the A(Ba) error ($\pm0.08$~dex) between 0.0035 to obtain a 23\% error of the s-process contribution. 
Now, the microturbulence ranges in Fig.~\ref{fig:vmic_var_4934} (horizontal axis) that correspond to this range of A(Ba) = $1.15 \pm 0.08$~dex are represented in  Fig.~\ref{fig:vcalib_REW} by the red error bars. 
The mean values of these errors are 0.1, 0.1, 0.2, and 0.2 km~s~$^{-1}$ in the uRGB, ocRGB, bRGB, and SG stages, respectively.
Therefore, these $v_{mic}$ errors are related to an error of 23\% error s-process determination. Table~\ref{tab:teff_errors} lists the percentage error related to $v_{mic} \pm 0.10$~km~s$^{-1}$ for every evolutionary stage.
The percentage errors in the TO stars were determined manually by changing $v_{mic}$ during the isotopic ratio determination.

The resonance line at 4934~\AA\ has often been avoided to derive isotopic ratios in the literature \citep[e.g. ][]{mashonkina2006A&A...456..313M, Gallagher2020A&A...634A..55G} because it is blended with a small Fe~I line
We provide estimates of the effects of the [Fe/H] errors on $\Sigma_{iso}$ to demonstrate that the contribution of the Fe line blending to the error budget is minor.
The errors are computed similarly to those calculated for \teff. We changed the true [Fe/H] by $\pm0.10$~dex, and we obtained the values in Table~\ref{tab:teff_errors}.

\section{Checking  the presence of stellar activity}
\label{sec:activity}

Our study includes two TO dwarf stars, which are at least 10 Gyr old according to isochrone fit of the cluster's sequence (see Fig.~\ref{fig:kiel}). At this age, no Ba excess is observed in dwarf field stars \citep[e.g. Fig.~1 in][]{reddy2017ApJ...845..151R}. 
Therefore, it is unlikely that the Ba abundances in our TO stars are influenced by the physical phenomena underlying the barium puzzle \citep[see, e.g.][]{dorazi2009ApJ...693L..31D, Baratella2020A&A...634A..34B}, and thus they can be considered reliable.
Regarding our RGB and SG stars, no measurements of Ba excess in similarly old stars at these evolutionary stages are available in the literature.

\begin{figure}
    \centering
    \includegraphics[width=1\linewidth]{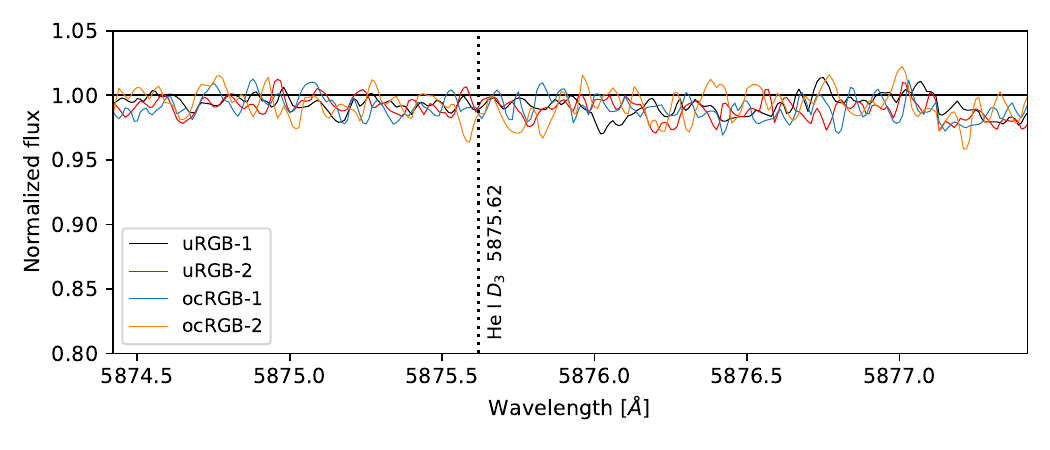}
    \caption{\tiny Spectra of our uRGB and ocRGB stars around the wavelength $\lambda$5875.62~\AA\ of the He~I $D_3$ line.
    The wavelength where the He~I $D_3$ line is indicated by the dashed line.
    }
    \label{fig:He_I}
\end{figure}

To confirm the absence of magnetic activity, we examinated the spectra of our sample stars searching for  the He~I $D_3$ line at $\lambda$5875.62~\AA, 
which is a chromospheric activity indicator; we show four spectra with the highest S/N in our sample in
Fig.~\ref{fig:He_I}. 
Not even the spectrum of the highest quality (star uRGB-1 with S/N = 191 at $\lambda$5875.62~\AA) shows any trace of the line. 
However, we can estimate the highest limit of its activity assuming that a line is masked by the spectral noise.
In this case, the line depth must be of 1\% at most, considering a 2$\sigma$ detection; roughly, the line would be of EW $\lesssim 2$~m\AA.
Assuming that the relation\footnote{We use Equation 1 of the paper.} between the He~I~$D_3$ index and the Ca~II~H and K index log $R'_{\mathrm{HK}}$ of \cite{reddy2017ApJ...845..151R} is suitable for metal-poor RGB stars, EW = 2~m$\AA$ is equivalent to log~$R'_{\mathrm{HK}} = -5$. This number indicates a base level of negligible chromospheric activity and null Ba excess (Figs.~9 and 10 in the paper), thus our RGB and SG stars should not present atypical enhancements due to chromospheric activity.

\section{The role of atomic diffusion}
\label{sec:diffusion}

In the literature, there has been some controversy on whether element abundance offsets between TO and RGB stars are proofs of atomic diffusion \citep[see e.g.][]{korn2008MmSAI..79..495K}. This is a topic of high relevance in fundamental physics because it could solve the \textit{cosmic lithium problem} \citep[e.g.][]{spite1982Natur.297..483S,ryan1999ApJ...523..654R, fields2011ARNPS..61...47F}.
Since our method is different to those used by diverse authors assessing the presence of {\it atomic diffusion} \citep[][among others]{korn2006Natur.442..657K,korn2007ApJ...671..402K, lind2008A&A...490..777L, nordlander2012ApJ...753...48N,gruyters2013A&A...555A..31G, gruyters2014A&A...567A..72G, souto2018ApJ...857...14S, souto2019ApJ...874...97S}
our results may be considered an independent test for the presence of diffusion in this cluster.
The argument against the diffusion hypothesis supports that the abundance offsets between the TO and the RGB are artifacts produced by spectroscopic methods.

\begin{figure}
    \centering
    \includegraphics[width=0.9\linewidth]{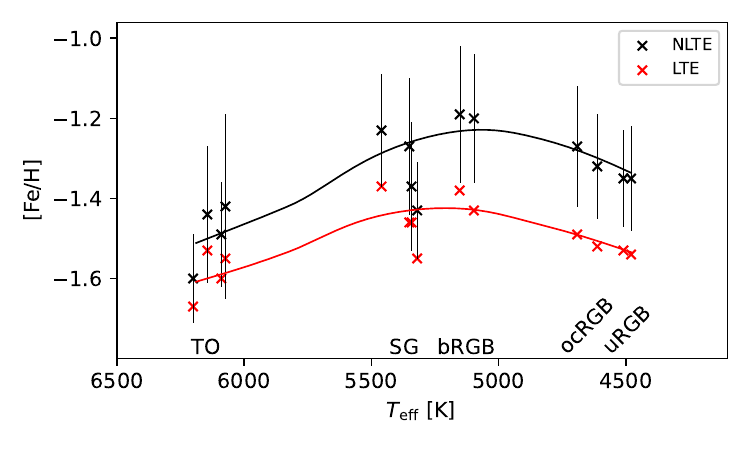}
    \caption{\tiny Metallicity as function of \teff. 
    Red and black crosses represent NLTE and LTE determinations, respectively.
    The errors of NLTE values are represented by the black bars.
    Red and black curves are corresponding LOWESS regressions (TO-2 neglected). 
    Evolutionary states are noted in the plot. 
    }
    \label{fig:diffusion}
\end{figure}

Figure~\ref{fig:diffusion} shows the trends of [Fe/H] as function of \teff\ and the evolutionary state, under LTE and NLTE.
Although there is an offset between the two sets of determinations, both show a systematic increase of [Fe/H] in the stars out of the TO.
The trend of our chosen NLTE [Fe/H] shows an increase of +0.25~dex at bRGB with respect to the TO (the TO-2 star was neglected in the LOWESS fit).
Compared with \cite{gruyters2013A&A...555A..31G}, who analyse the same cluster, under 1D LTE, we observe a similar diffusion effect.
Under 1D~NLTE, we measure an effect that is approximately +0.15~dex higher than the observed [Fe/H], and +0.10 to +0.15~dex above theoretical predictions.

In the following, we investigate whether the [Fe/H] offset between the TO and RGB stars shown in Fig.~\ref{fig:diffusion} may result from our spectroscopic analysis method. 
First, we applied the \ion{Fe}{ii} 3D~NLTE corrections from \cite{amarsi2016MNRAS.463.1518A} to the 1D~NLTE [Fe/H] values in Fig.~\ref{fig:diffusion}, obtaining an increase by +0.10~dex for both TO and RGB stars. Therefore, 3D~NLTE adjustments cannot reconcile the observed [Fe/H] offset.
We also tested the \teff\ scale, as the impact of \logg\ changes in [Fe/H] are usually negligible\footnote{The highest impact of \logg\ is the SG branch, where changes of $\pm0.3$~dex induce [Fe/H] $\mp0.1$~dex, approximately.}.
Assuming the 1D~NLTE framework, reconciling the [Fe/H] offset would require adjusting the TO stars’ 
\teff upward by roughly 250~K, or lowering the RGB stars’ \teff by a similar amount.
Under 3D~NLTE H$\alpha$ modelling \citep{giribaldi2021A&A...650A.194G, giribaldi2023A&A...679A.110G}, the \teff\ scale shifts by at most 75~K. Therefore, H$\alpha$‑based 3D~NLTE \teff\ determinations cannot account for the 250~K adjustment required.

Finally, in \cite{giribaldi2021A&A...650A.194G}, we have shown that residual instrument patterns in UVES spectra may affect H$\alpha$ \teff, however no deviation larger than 80~K was observed (Fig.~6 in the paper).
Therefore, given current evidence, the diffusion pattern shown in the figure is most likely authentic.

\section{Extra Figures}

\begin{figure}[!h]
    \centering
    \includegraphics[width=0.9\linewidth]{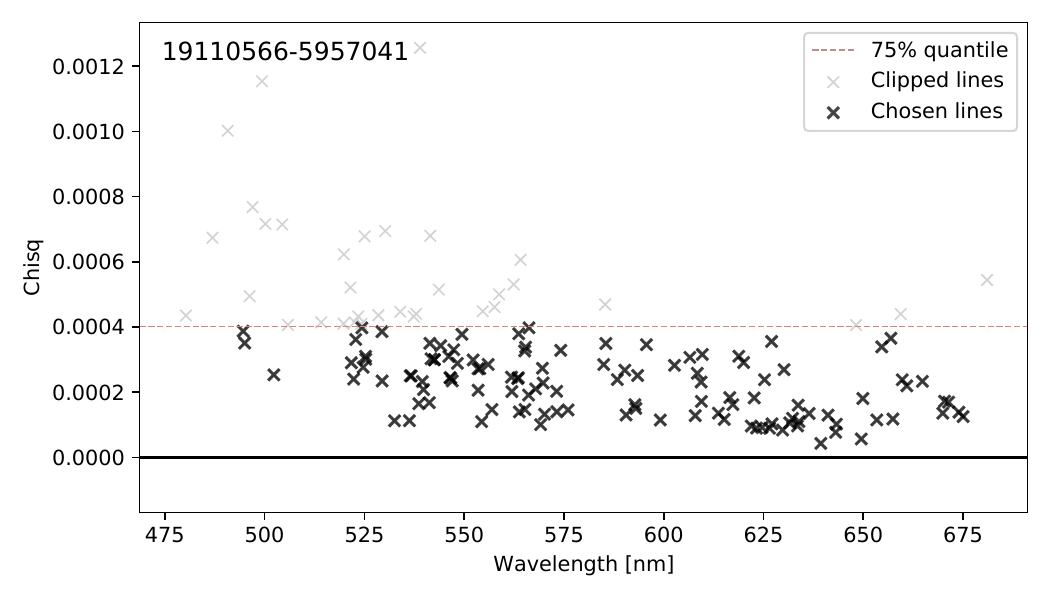}
    
    \caption{\tiny $\chi^2$ of the Gaussian fits of Fe lines as function of the wavelength.
    This plot corresponds to the star bRGB-1 or 19110566-5957041.
    Dark crosses represent accepted lines, whereas gray crosses represent clipped lines. The dashed red line separates the 75\% quantile.
    }
    \label{fig:clipping_lines}
\end{figure}

\begin{figure}[!h]
    \centering
    \includegraphics[width=0.48\linewidth]{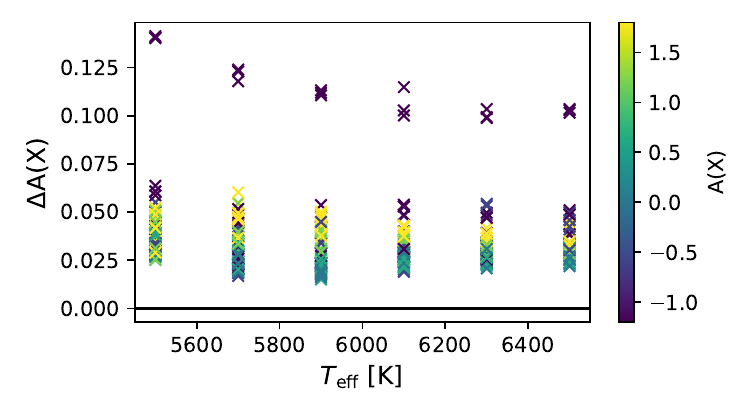}
    \includegraphics[width=0.48\linewidth]{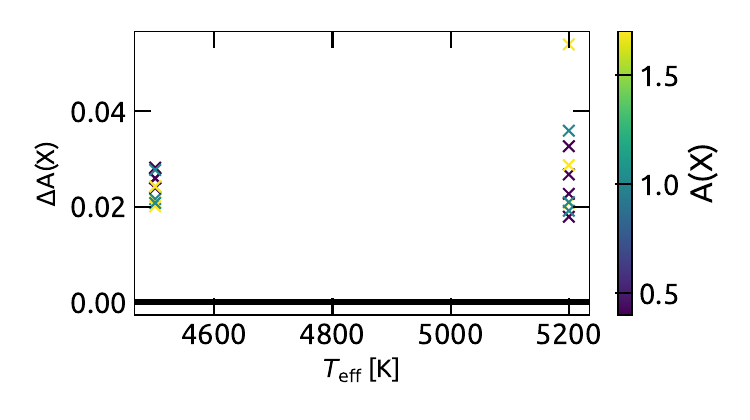}
    \includegraphics[width=0.48\linewidth]{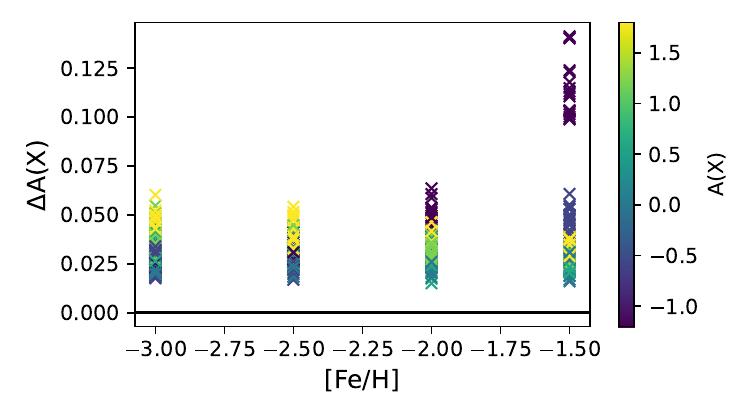}
    \includegraphics[width=0.48\linewidth]{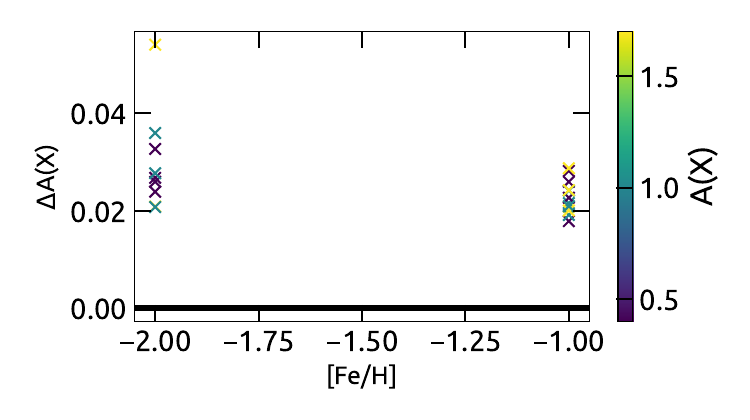}
    
    \caption{\tiny A(Ba) grid of offsets induced by    
    \teff\ (+40~K) offsets when derived from the line 6141~\AA.
    The left and right columns display the grids of TO and RGB stars, respectively. 
    The horizontal line represents no offset. "X" represents the element, Ba in this case.}
    \label{fig:Ba_dif}
\end{figure}

\begin{figure}
    \centering
    \includegraphics[width=0.8\linewidth]{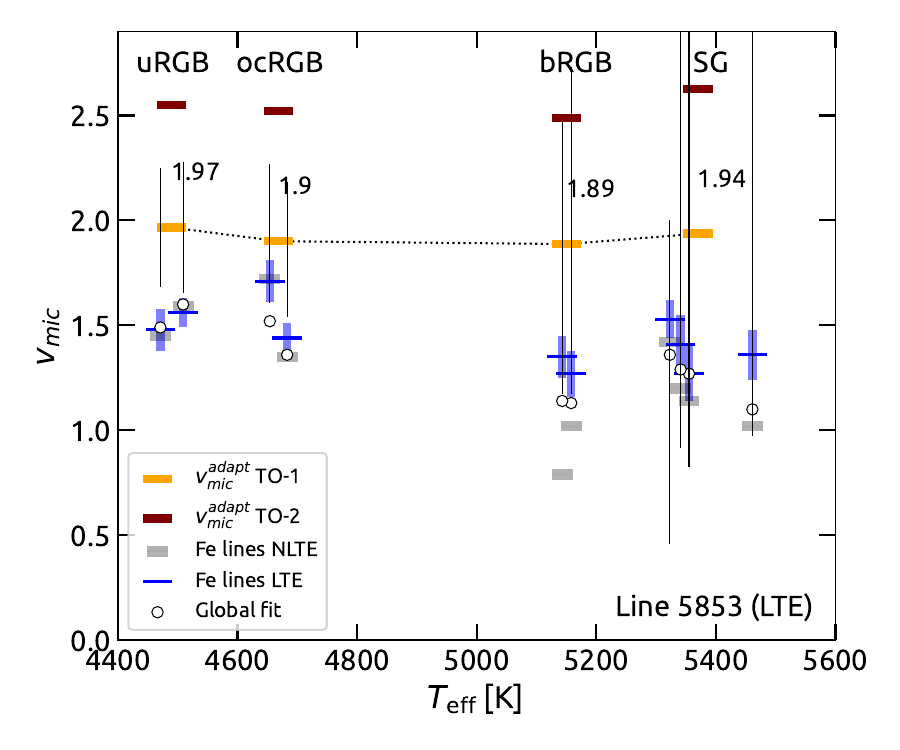}
    \caption{\tiny Same as left panel in Fig.~\ref{fig:vmic_teff}, but its values are from A(Ba) = 0.89~dex, instead of A(Ba) = 1.13~dex. 
    }
    \label{fig:vmic_logg_5853_no_cor}
\end{figure}

\begin{figure}
    \centering
    \includegraphics[width=0.48\linewidth]{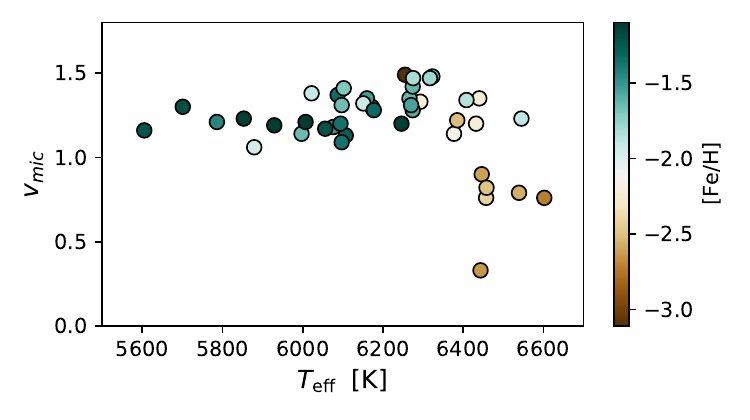}
    \includegraphics[width=0.48\linewidth]{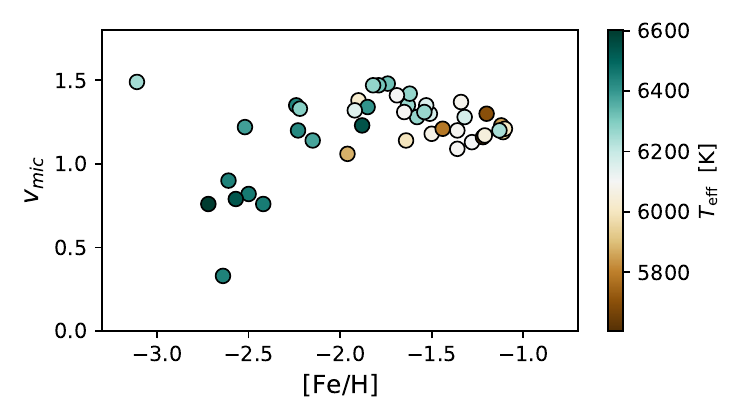}
    \includegraphics[width=0.48\linewidth]{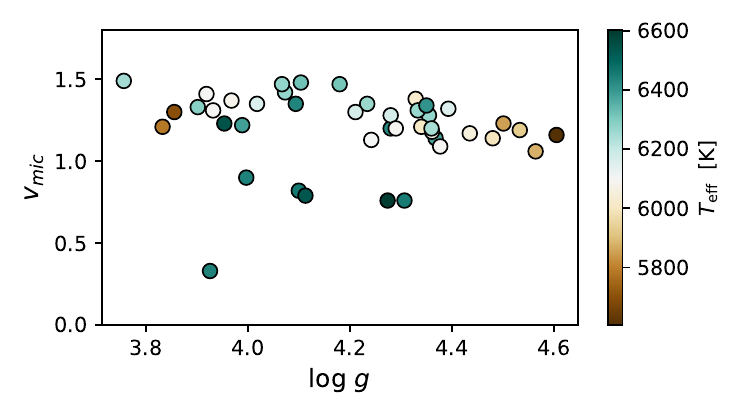}
    \caption{\tiny Microturbulence of the \titan~I dwarf sample. Values are plotted as functions of the atmospheric parameters following the colour-coding of the pallets.}
    \label{fig:vmic_titans}
\end{figure}

\begin{figure*}[t]
    \centering
    \includegraphics[width=0.3\linewidth]{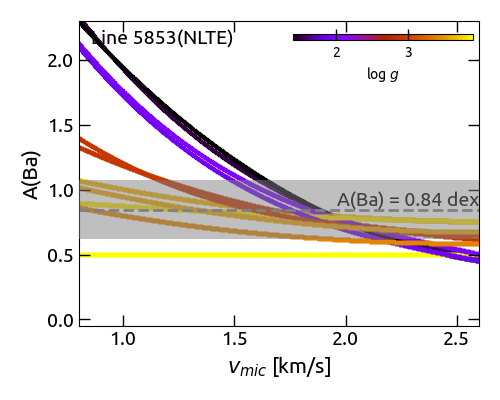}
    \includegraphics[width=0.3\linewidth]{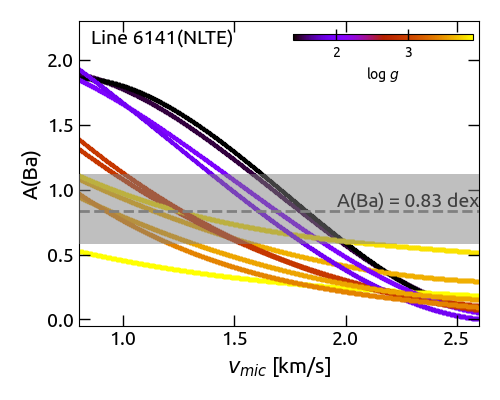}
    \includegraphics[width=0.3\linewidth]{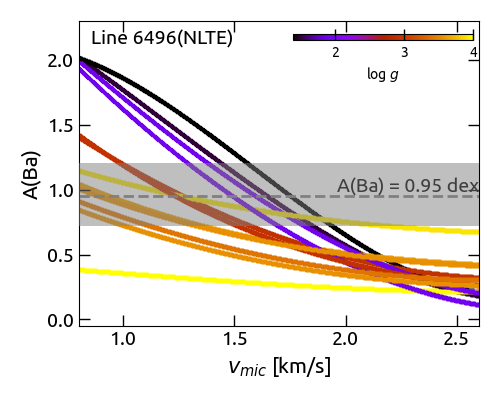}
    \caption{\tiny Similar to Fig.~\ref{fig:vmic_var}, but under NLTE. A(Ba) was derived using the Ba model atom in \citet{gerber2023} and the departure coefficients of \citep{Gallagher2020A&A...634A..55G}.
    }
    \label{fig:vmic_var_NLTE}
\end{figure*}

\end{appendix}

\end{document}